\documentstyle[12pt]{article}
\begin{document}
{\rm

\title{ Nuclear Transparency \\ in \\ 
          a Relativistic Quark Model  }
\author{ Tetsu IWAMA, Akihisa KOHAMA\thanks{Correspondences to:~
kohama@tkynt2.phys.s.u-tokyo.ac.jp}
 and Koichi YAZAKI  \\\\
           Department of Physics, Faculty of Science,\\
           University of Tokyo,\\
           7-3-1 Hongo, Bunkyo-ku, Tokyo 113, JAPAN }

\maketitle

\begin{abstract}
We examine the nuclear transparency for the quasi-elastic ($e, e'p$) 
process at large momentum transfers in a relativistic
quantum-mechanical model for the internal structure of the proton,  
using a relativistic harmonic oscillator model. 
A proton in a nuclear target is struck by the incident electron 
and then propagates through the residual nucleus 
suffering from soft interactions with other nucleons. 
We call the proton ``dynamical" 
when we take into account of internal excitations, 
and ``inert" when we freeze it to the ground state. 

When the dynamical proton is struck with a hard 
(large-momentum transfer) interaction, it shrinks, 
{\it i.e.}, small-sized configuration dominates the process. 
It then travels through nuclear medium 
as a time-dependent mixture of 
intrinsic excited states and thus changing its size. 
Its absorption due to the soft interactions with nuclear medium 
depends on its transverse-size. 
Since the nuclear transparency is a measure of the absorption strength, 
we calculate it in our model for the dynamical case, 
and compare the results with those for the inert case. 
The effect of the internal dynamics is observed,  
which is in accord with the idea of the ``color transparency". 
We also compare our results with the experimental data 
in regard of $q^2$-dependence as well as $A$-dependence, 
and find that the $A$-dependence may reveal 
the color-transparency effect more clearly. 

Similar effects of the internal dynamics 
in the other semi-exclusive hard processes are briefly discussed. 
\end{abstract}

\newpage
\setcounter{section}{0}
\setcounter{equation}{0}
\section{Introduction}
\label{sec-intro}

A phenomenon ``color transparency" was predicted 
by Brodsky and Mueller in the early 80's
\cite{Brodsky1}, \cite{Brodsky2} 
as a candidate of observing the effects of internal dynamics 
of the proton in high-energy nuclear reactions. 
It was speculated 
that the initial- and/or final-state interactions of a proton 
involved in a high-momentum transfer reaction 
with a nuclear target would be suppressed, 
and that the nuclear medium would look transparent
\footnote{For recent reviews, 
see Refs.~\cite{NNN:Kik}, \cite{Fra:Ann}, and \cite{Ral:Rep}}.

The idea of the color transparency can be explained 
in the following way. 
We call the proton ``dynamical" 
when we take into account of internal excitations, 
and ``inert" when we freeze it to the ground state: 
\begin{itemize}
\item 
Let us consider exclusive processes, such as 
electron-proton or proton-proton elastic scatterings 
at large-momentum transfers. 
If we take account of the internal structure of the proton 
consisting of quarks and gluons, 
the above processes are dominated by small-sized configurations 
with the minimum number of constituents 
since the transferred large-momentum must be shared 
by all the constituents in the exclusive processes. 
\item
In Quantum Chromodynamics (QCD), 
the interactions are mediated by the color field, 
but it does not couple with point-like color neutral objects, 
because the coupling is proportional to the color charges. 
Therefore, if the above exclusive processes occur in nuclear medium, 
the proton shortly before or after the hard process interacts weakly with 
the surrounding nucleons. 
We thus expect that the initial- and/or final-state interactions 
of a dynamical proton involved in such a hard semi-exclusive process 
with a nuclear target 
will be weaker than those expected for an inert proton. 
\end{itemize}

The possibility of observing the color transparency 
in the quasi-elastic processes such as ($e, e'p$) and ($p, 2p$), 
has been discussed by many authors \cite{Farrar}-\cite{Kohama1}. 
A novel feature of the present work is 
the use of a relativistic model for the internal dynamics, 
which will enable us to examine the prescriptions of 
incorpolating relativistic effects in the previous discussions 
and tells us their importance in a transparent way.

%%%% Definition
We first introduce a quantity called ``nuclear transparency", 
$T(|{\bf k}|, \hat{{\bf k'}})$,   
for a quasi-elastic process, $(h, h'p)$, on a nuclear target, $A$. 
\begin{equation}
  T(|{\bf k}|, \hat{{\bf k'}}) = \frac{1}{Z} \frac{1}{A({\bf k})} 
         \frac{d\sigma_{hA}}{d\Omega_{k'}} /
         \frac{d\sigma_{hp}}{d\Omega_{k'}}, 
\label{eq-def}
\end{equation}
where ${\bf k}$ and ${\bf k'}$ are 
the momenta of the incident and the scattered particles, 
$h$ and $h'$, respectively, 
and $Z$ is the atomic number of the target nucleus. 
Here $A({\bf k})$ is the Fermi-motion averaging factor. 
The denominator is the $h p$ elastic differential cross section, 
while the quasi-elastic differential cross section in
the numerator is defined by
\begin{equation}
  \frac{d\sigma_{hA}}{d\Omega_{k'}} = 
  \int |{\bf k'}|^{2} \, d|{\bf k'}| d{\bf p} \;
  \frac{d\sigma_{hA}}{d{\bf k'} d{\bf p}},  
\end{equation}
where ${\bf p}$ is the recoil momentum of the struck proton. 

Our definition of the nuclear transparency is somewhat unusual in that
the quasi-elastic cross section is fully integrated over the proton
recoil momentum and the energy of the scattered particle. 
In fact, 
it does not correspond to what have been measured experimentally. 

%%% Experiments
There have been two experimental studies of the color transparency,
one with $(p, 2p)$ at BNL \cite{Carroll}
and the other with $(e, e'p)$ at SLAC \cite{Makins}, \cite{ONe:PRL}.
In either case, the measured quantity is not exactly the transparency, 
$T(|{\bf k}|, \hat{{\bf k'}})$, defined here, 
because of the limited kinematical
region covered by the experiment.
We nonetheless employ this definition for our discussion,
since it allows us the most reliable theoretical treatment, 
on the one hand, 
and the required measurements are accessible in future experiments, 
such as the Jefferson Lab. (former CEBAF),
being not too far from the actually performed ones, on the other hand. 
The nuclear transparency defined in this way is a function of two
variables, {\it i.e.}, the magnitude of the incident momentum, $|{\bf k}|$, 
and the scattering angle, $\theta ({\bf k}\wedge{\bf k'})$.

In the following, we deal mainly with the simpler case 
of electron scattering, $(e, e'p)$. 
In this case, since the initial- and final-state
interaction of the electron can be neglected, 
the transparency depends only on the magnitude 
of the average three-momentum transfer
determined by the above two variables through the free kinematics, 
{\it i.e.}, 
\begin{equation}
  T(|{\bf k}|, \hat{{\bf k'}}) = T(|\bar{{\bf q}}|).
\end{equation}
Here $|\bar{{\bf q}}|$ is determined by
\begin{equation}
   |{\bf k}| + M = |\bar{{\bf k'}}| 
            + \sqrt{\bar{{\bf q}}^{2} + M^{2}}, ~~~
  \bar{{\bf q}}^{2} = {\bf k}^{2} + \bar{{\bf k'}}^{2} 
           - 2 |{\bf k}| |\bar{{\bf k'}}| \cos{\theta}, 
\end{equation}
where the electron mass is ignored. 
$M$ is a nucleon mass, and 
$\bar{{\bf k'}}$ is the average recoil momentum.

In $(e, e'p)$, 
the Fermi-motion averaging factor in eq.~(\ref{eq-def}), 
$A({\bf k}) \simeq 1.05$, is 
almost independent of ${\bf k}$ \cite{Seki1}.
A characteristic feature of the electron quasi-elastic process 
is that the relevant soft interaction operates only 
on the emitted proton whose momentum is approximately equal to 
the transferred momentum. 
The proton is thus initially struck in the longitudinal direction 
while the soft interaction is sensitive to its transverse size. 
The longitudinal-transverse correlation is therefore necessary 
for the color transparency to be observed in the $(e, e'p)$ process. 
We should keep this in our mind in constructing the model. 

In this work, 
we use the result of the Glauber Impulse Approximation (GIA)
\cite{Seki1} under the zero-range-no-recoil (ZRNR) approximation 
as a reference frame representing the case
of no internal dynamics for the proton.  
Then we introduce a model for the internal dynamics of the proton 
to see the effect of the color transparency.
In order to describe the internal dynamics, 
we use a relativistic quark model. 
The reason why we treat this problem in a covariant way is that,
if we treat this problem non-relativistically,
the internal velocity of a quark can be arbitrary large,
being proportional to its momentum, 
and the dynamical proton could expand too fast 
suppressing the color transparency effect.

Two of the present authors calculated
the nuclear transparency with a model emphasizing the breathing mode 
of the nucleon (the $b$-model) \cite{Kohama1}.
They included the time dilation factor 
to incorpolate the relativistic effects 
as had been done in the other approaches, 
but there has been no justification for such a prescription. 
We will examine here its validity in a fully relativistic model.

We calculate the nuclear transparency for the dynamical proton 
in the present model, 
and compare it with that for the inert proton. 
The comparison between the two cases 
is expected to show us the effect of the color transparency.

%% Contents
The contents of this paper are organized as follows: 
The expression based on the Glauber approach, which does not take
account of the internal dynamics and can thus be used as a reference
frame for observing the ``color transparency", 
is reviewed in sec.~\ref{sec-gia}.
A formalism of incorporating the internal dynamics in calculating 
the nuclear transparency is briefly described in sec.~\ref{sec-timedep}. 
A relativistic harmonic oscillator model for the proton is introduced 
in sec.~\ref{sec-model}. 
We review the original version in subsec.~\ref{sec-harmonic}, 
and introduce some modification so as to incorporate 
the longitudinal-transverse correlation in subsec.~\ref{sec-ltcor}.  
The proton survival amplitude is given in subsec.~\ref{sec-amp}. 
The numerical results and discussion are given 
in sec.~\ref{sec-nume}. 
The importance of the relativity 
and the correlation in the internal dynamics
will also be clarified there. 
The summary and the conclusion are given in sec.~\ref{sec-summary}.

\newpage
\setcounter{equation}{0}
\section{Glauber Approach}
\label{sec-gia}

Here we quote the result of
the Glauber Impulse Approximation (GIA) \cite{Seki1}. 
It is based on the Glauber multiple-scattering (not eikonal) theory 
\cite{Kerman}, \cite{Glauber} for the proton-nucleus 
final-state interaction under zero-range-no-recoil (ZRNR) approximation
which neglects the target nuclear recoil and 
the finite $N N$-interaction range. 
The nuclear transparency for $(e, e'p)$ in this approximation is given by
\begin{equation}
  T(|{\bf q}|) = \int d{\bf r} \; \rho({\bf r}) \; 
                           P^{(-)}({\bf q}; {\bf r}),
\label{T-P}
\end{equation}
where, without the nuclear correlation,  
the proton survival probability, 
$P^{(-)}({\bf q}; {\bf r})$, can be expressed as
\begin{equation}
P^{(-)}({\bf q}; {\bf r}) = \exp \{-(A - 1) \; 
            \sigma^r_{{\rm NN}}(|{\bf q}|)
                     \int_{z}^{\infty} \! dz^\prime \;
                     \rho({\bf b}, z^{\prime}) \}, ~~
     z \parallel {\bf q}, ~~    {\bf r} = ({\bf b}, z). 
\label{P-def}
\end{equation}
Here $\rho({\bf r})$ is the nuclear density, 
and $P^{(-)}({\bf q}; {\bf r})$ includes 
the effects of the final-state interactions
due to multiple scattering.
${\bf r}$ indicates the point where the proton has been struck,
and the path of integral in eq.~(\ref{P-def}) is taken to be along
the classical path of the struck proton.
$\sigma^r_{\rm NN}(|{\bf q}|)$ 
$(= \sigma^{total}_{\rm NN}(|{\bf q}|) 
- \sigma^{elastic}_{\rm NN}(|{\bf q}|))$
is the proton-nucleon reaction cross section 
at the incident momentum, $|{\bf q}|$. 

The expressions, eqs.~(\ref{T-P}) and (\ref{P-def}), have
a simple interpretation.
The proton is struck at the point ${\bf r}$ and propagates
along the path which is taken to be in the direction of $z$-axis,
disappearing at the rate,
$v_q \cdot \sigma^r_{{\rm NN}}(|{\bf q}|) 
\cdot (A - 1) \cdot \rho({\bf r})$,
where $v_q$ is a proton velocity.

Introducing the time, $t$, by $z = v_q \; t$, 
one can regard the propagation
as the time development of the struck proton in the nuclear medium.
We will take such a time dependent picture in the following.

For the nuclear correlation 
the detailed discussions can be found in 
Refs.~\cite{Seki1}, \cite{Niko:PLB1}, \cite{Sek:PLB}, \cite{Rin:NPA}.

\newpage
\setcounter{equation}{0}
\section{Time Dependent Description}
\label{sec-timedep}

In this section, 
we explain the time dependent description 
of the internal dynamics \cite{proc}, 
and introduce the form factor and the survival amplitude. 
We consider the case where the incident electron hits a proton
in the target nucleus at the time, $t = 0$, 
and follow the time development of the struck proton 
as it travels in the medium.

We describe an eigenstate of a free dynamical proton as 
$|N; {\bf P} \rangle$, 
where $N$ and ${\bf P}$ here are a set of quantum numbers 
of internal excitations 
and the center-of-mass spatial momentum of the proton, respectively. 
The internal ground state corresponding 
to the ordinary proton is expressed by $N = 0$.
The elastic form factor, $F_{{\rm ep}}(q^{2})$, is given by
\begin{equation}
  F_{{\rm ep}}(q^{2}) 
  = \langle 0; {\bf P}_{f} | \hat{O}(q) |0; {\bf P}_{i} \rangle,
\label{form0}
\end{equation}
where $\hat{O}(q)$ is the hard interaction operator 
with the four-momentum transfer, $q_{\mu}$ $= P_{f, \mu} - P_{i, \mu}$. 
$P_{i, \mu}$ $(P_{f, \mu})$ is the four-momentum of the proton 
in its initial (final) state.

Then, we introduce the full hamiltonian, $\hat{H}$,  
\begin{equation}
  \hat{H} = \hat{H}_{0} + \hat{V}, 
\end{equation}
where $\hat{H}_{0}$ is the hamiltonian of the free proton 
whth internal dynamics, 
and $\hat{V}$ is the interaction 
between the struck proton and the surrounding nuclear medium. 
Thus, the time development of the dynamical proton 
as it travels through the nuclear medium 
is given by $\exp(-{\it i} \hat{H} t)$.

The desired matrix element is the probability amplitude for the system 
to be in the ground state of the physical proton 
after the interaction with the medium 
for a period of time, $t$, which is given by
\footnote{The index `$(D)$' represents the dynamical proton.}
\begin{equation}
  M^{(D)}_{{\rm eA}} (q^2; t) = \langle 0; {\bf P}_{f} | 
        e^{-{\it i} \hat{H} t} \hat{O}(q) |0; {\bf P}_{i} \rangle.
\label{matrix0}
\end{equation}
Note that $M^{(D)}_{{\rm eA}}(q^{2}; t = 0)$ coincides
with the form factor, $F_{{\rm ep}}(q^{2})$, eq.~(\ref{form0}).
For later convenience, 
we define a ratio, $R(q^2; t)$, by   
\begin{equation}
  R(q^2; t) \equiv M^{(D)}_{{\rm eA}}(q^{2}; t)
                 / (e^{-{\it i} \hat{H}_0 t} \; F_{{\rm ep}}(q^{2}) ).  
\label{ratio1}	
\end{equation}
The survival probability, $P^{(-)}({\bf q}; {\bf r})$, 
which gives the nuclear transparency through eq.~(\ref{T-P}), 
is then given by 
\begin{equation}
  P^{(-)}({\bf q}; {\bf r}) = |R(q^{2}; t({\bf r}))|^2, 
\label{P-}	
\end{equation}
where $t$ is the propagation time for the struck proton 
from the point, ${\bf r}$, to the nuclear surface. 
The effect of smooth nuclear surface is
approximately included by taking $t$ as
\begin{equation}
  t({\bf r}) = \frac{1}{v_{P_f} \; \rho_0}
             \int_z^\infty dz^\prime \; \rho({\bf b}, z^{\prime}), 
\label{t-r}
\end{equation}
where $v_{P_f}$ is the velocity of the struck proton, 
and $\rho_0$ is the value of $\rho({\bf r})$ at the origin.
The expression, eq.~(\ref{t-r}), is justified 
if the interaction strength is weak 
and the modification of the time evolution is proportional 
to the density. 

We also note the following property of the amplitude, 
$M^{(D)}_{{\rm eA}}(q^{2}; t)$. 
Assuming that $\hat{V} = 0$, {\it i.e.}, no interaction, 
the matrix element becomes
\begin{eqnarray}
 M^{(D)}_{{\rm eA}}(q^{2}; t) 
     =  \langle 0; {\bf P}_{f}| e^{-{\it i} \hat{H}_0 t}\;  
        \hat{O}(q)   |0; {\bf P}_{i} \rangle  
    = e^{-{\it i} E_{P_f} t} \; F_{{\rm ep}}(q^2), 
\label{matrix-0}
\end{eqnarray}
where we have used the fact that $|0; {\bf P} \rangle$
is an eigenstate of $\hat{H}_0$, {\it i.e.}, 
\begin{eqnarray}
  \hat{H}_0 \; |0; {\bf P} \rangle =  E_P \; |0; {\bf P} \rangle, 
\end{eqnarray}
where $E_P$ $= \sqrt{M^2 + {\bf P}^2}$, 
and $M$ denotes the proton mass. 

Thus, in the absence of the interaction with the nuclear medium, 
$|M^{(D)}_{{\rm eA}}(q^{2}; t)|^2$ becomes a mere squared 
form factor, $|F_{{\rm ep}}(q^2)|^2$,  
and $P^{(-)}({\bf q}; {\bf r})$ becomes unity, 
giving the complete transparency, $T(q) = 1$.
The deviations from unity of $P^{(-)}({\bf q}; {\bf r})$ 
and of $T(q)$ reflect the effects of the interaction, $\hat{V}$.

Next we consider the matrix element, 
$M^{(I)}_{{\rm eA}}(q^{2}; t)$, for an inert proton
\footnote{The index `$(I)$' represents the inert proton.}.
The inert proton is never excited, staying only in the ground state, 
and $M^{(I)}_{{\rm eA}}(q^{2}; t)$ becomes
\begin{equation}
  M^{(I)}_{{\rm eA}}(q^{2}; t) 
=  e^{-{\it i} \langle  \hat{H}  \rangle t} \; F_{{\rm ep}}(q^2) 
=  e^{-{\it i} (E_{P_f} + \langle  \hat{V}  \rangle ) t} \; 
       F_{{\rm ep}}(q^2).
\label{matrix_inert}
\end{equation}
Here we have defined 
\begin{equation}
  \langle  \hat{O}  \rangle 
  \equiv \langle 0; {\bf P}_{f}| \hat{O} |0; {\bf P}_{f} \rangle.
\end{equation}
Following eq.~(\ref{ratio1}), 
we define a similar quantity for the inert proton as 
\begin{eqnarray}
  R_{I}(q^2; t) 
  &\equiv& M^{(I)}_{{\rm eA}}(q^{2}; t)
                 / (e^{-{\it i} \hat{H}_0 t} \; F_{{\rm ep}}(q^{2}) )
\nonumber \\
  &=& e^{-{\it i} \langle  \hat{V}  \rangle t}. 
\label{ratio2}	
\end{eqnarray}
The survival probability, $P^{(-)}({\bf q}; {\bf r})$, is then obtained as
\begin{eqnarray}
P^{(-)}({\bf q}; {\bf r}) 
   &=& |R_{I}(q^{2}; t({\bf r}))|^2   \nonumber\\
   &=& |\exp \{-{\it i} 2 \langle \hat{V} \rangle t({\bf r}) \}| 
        = \exp \{2~{\rm Im}~\langle \hat{V} \rangle  t({\bf r}) \}. 
\label{P-inert}
\end{eqnarray}

The expression~(\ref{P-inert}) becomes identical 
with the Glauber expression~(\ref{P-def}) if we take 
\begin{equation}
 - 2~{\rm Im}~\langle 0; {\bf P} | \hat{V} |0; {\bf P} \rangle
 = \frac{v_P}{\lambda}
  = \sigma^r_{{\rm NN}}(|{\bf P}|)  \; \rho_{{\rm NM}} \; v_P.
\label{c-inert}
\end{equation}
Here $\sigma^r_{{\rm NN}}(|{\bf P}|)$, $\rho_{{\rm NM}}$ and $\lambda$ are
the proton-nucleon reaction cross section, 
the nuclear matter density and
the mean-free path of the proton in the nuclear matter, respectively, 
and we note that $\rho_{{\rm NM}}$ $\simeq (A - 1) \rho_0$ 
for our normalization of $\rho({\bf r})$. 
Later we adopt the relation~(\ref{c-inert}) 
in order to determine the strength of the interaction operator, $\hat{V}$.

\newpage
\setcounter{equation}{0}
\section{Formulation of the Model}   
\label{sec-model}

In this section we briefly review the original version 
of the relativistic harmonic oscillator model 
\cite{Takabayashi}, \cite{Namiki1}, \cite{FKR:PRD}, 
and then modify it for incorporating 
the longitudinal-transverse correlation
which we introduce to describe the color transparency. 

The first attempt to include the internal dynamics of 
the proton was made by Farrar {\it et al}. \cite{Farrar}, 
using the model of the classically-expanding proton. 
Many other authors made descriptions 
of the internal structure of the struck proton 
based on tha hadronic basis \cite{Jennings}, \cite{NNN:NPA}, and 
on the non-relativistic quark model \cite{BZK:PLB}, \cite{Kohama1}. 
Nobody has pointed out the importance 
of the relativity for the internal dynamics so far. 
This is the point that we will clarify in this work. 

\subsection{Relativistic Harmonic Oscillator Model}	
\label{sec-harmonic}

Now we introduce a relativistic quark model for the proton.
A relativistic 4-dimensional harmonic oscillator model is
very appropriate for our purpose, 
since it reproduces the elastic form factor 
as well as the excitation spectrum represented 
by the linearly rising Chew-Frautschi trajectory \cite{Takabayashi}.

In this formulation the color-singlet three-quark system obeys 
a relativistic wave equation,
\begin{eqnarray}
   (P^{2} - \hat{M}^{2})|\Psi; {\bf P} \rangle = 0.
\end{eqnarray}
Here $\hat{M}$ is the mass operator, and carries the information 
of the internal dynamics. 
An internal state of the proton is governed by an eigenvalue equation, 
\begin{eqnarray}
   \hat{M}^{2} \; |\phi_{n}; {\bf P} \rangle 
 = M_{n}^{2}   \; |\phi_{n}; {\bf P} \rangle. 
\label{eq:rqm1}
\end{eqnarray}
Here we have introduced the notation, $\phi_{n}$, 
for specifying states in free space, 
while we use $\phi_{I, n}$ for those interacting with nuclear medium  
which we introduce later. 
The ground state, $\phi_0$, corresponds to the physical proton. 

The mass operator for the three-quark system 
in the relativistic 4-dimensional
harmonic oscillator model is given by \cite{Namiki1}
\begin{equation}
  - \hat{M}^2 = \eta \left( \hat{p}_{r}^{2} + \hat{p}_{s}^{2} 
        + \alpha^2 (\hat{r}^2 + \hat{s}^2) \right)  + C,
\label{oscillator}
\end{equation}
where $r^2$ $= r_{0}^{2} - {\bf r}^2$. 
$\alpha$ is a size parameter and is chosen so as to reproduce 
the observed form factor. 
$\eta$ and $C$ are then adjusted to reproduce the proton mass 
of 940~[MeV] and the Roper resonance of 1440~[MeV]. 
These parameters are given in subsec.~\ref{sec-para}.

We have rearranged the coordinate four-vectors for three quarks, 
$\hat{x}_{1, \mu}$, $\hat{x}_{2, \mu}$, 
and $\hat{x}_{3, \mu}$, as follows:
\begin{eqnarray}
\hat{X}_{\mu} &=& {1 \over 3} 
      (\hat{x}_{1, \mu} + \hat{x}_{2, \mu} + \hat{x}_{3, \mu}),  \\
\hat{r}_{\mu} &=& {1 \over \sqrt{6}} 
                 (\hat{x}_{2, \mu} - \hat{x}_{3, \mu}), ~~~
\hat{s}_{\mu}  = {1 \over 3 \sqrt{2}} 
      (-2 \hat{x}_{1, \mu} + \hat{x}_{2, \mu} + \hat{x}_{3, \mu}), 
\label{henkan1}
\end{eqnarray}
Here $\hat{X}_\mu$ is the center-of-mass coordinate, and
$\hat{r}_\mu$ and $\hat{s}_\mu$ are the two independent relative coordinates
which represents internal structure.
The conjugate momenta are 
\begin{eqnarray}
\hat{P}_{\mu} &=& 
        \hat{p}_{1, \mu} + \hat{p}_{2, \mu} + \hat{p}_{3, \mu}, \\
\hat{p}_{r, \mu} &=& 
\sqrt{\frac{3}{2}} \; (\hat{p}_{2, \mu} - \hat{p}_{3, \mu}), ~~~
\hat{p}_{s, \mu} = \frac{1}{\sqrt{2}} \;
  (-2 \hat{p}_{1, \mu} + \hat{p}_{2, \mu} + \hat{p}_{3, \mu}).
\end{eqnarray}

The eigenvalues for time-axis are negative, 
and we call the highest-eigenvalue state the ``ground state". 
As was discussed by Takabayashi \cite{Takabayashi},
we need the additional condition;
\begin{equation}
 \hat{P} \cdot (-{\it i} \hat{p}_{r} + \alpha \hat{r}) \; 
 \Psi(r, s; {\bf P}) 
= \hat{P} \cdot (-{\it i} \hat{p}_{s} + \alpha \hat{r}) \; 
\Psi(r, s; {\bf P}) = 0, 
\label{condition}
\end{equation}
to get the spectrum bounded from below. 
In the rest frame, the condition, eq.~(\ref{condition}), 
restricts the eigenstate for time-axis to the ground state. 

For a free proton at rest, {\it i.e.}, ${\bf P} = {\bf 0}$, 
the ground-state wave function of the eigenvalue equation, 
eq.~(\ref{eq:rqm1}), 
under the condition, eq.~(\ref{condition}), is 
\begin{equation}
  \phi_0(r, s; {\bf 0}) 
  = \left( \frac{\alpha}{\pi} \right)^2 \;
 \exp \left[ - \frac{\alpha}{2} \; 
          (r_0^2 + {\bf r}^2 + s_0^2 + {\bf s}^2) \right].
\label{reststate0}
\end{equation}
Then, for a moving free proton which has a spatial momentum, ${\bf P}$, 
the eigenfunction, $\phi_0(r, s; {\bf P})$, should be 
the Lorentz-boosted form of eq.~(\ref{reststate0}),
\begin{equation}
  \phi_0(r, s; {\bf P}) = \left( \frac{\alpha}{\pi} \right)^2 \!
    \exp \left[ \frac{\alpha}{2} \; \{ r^2 + s^2 
         - \frac{2}{M_0^2} (P \cdot r)^2
         - \frac{2}{M_0^2} (P \cdot r)^2 \} \right], 
\label{movestate0a}
\end{equation}
where $M_0$ is the lowest eigenvalue of the mass operator, 
eq.~(\ref{oscillator}). 
This form is manifestly covariant. 
Since we are interested in the case of 
$P_{i, \mu}$ $=(M_0, {\bf 0})$, and  
$q_{\mu}$ $= (q_0, {\bf 0}_{\perp}, q_3)$,  
eq.~(\ref{movestate0a}) becomes
\begin{eqnarray}
 \phi_0(r, s; {\bf q})  
 &= & \left( \frac{\alpha}{\pi} \right)^2 \; 
   \exp [ \frac{\alpha}{2} \; \{ r^2 + s^2  
            - \frac{2}{M_0^2} 
                \left((M_0 + q_0) r_0 - q_3 r_3 \right)^2 \nonumber \\
 &{}& ~~~~~~~~~~  - \frac{2}{M_0^2} 
                \left((M_0 + q_0) s_0 - q_3 s_3 \right)^2 \} ].
\label{movestate0}
\end{eqnarray}

Next, we consider the proton moving through the nucleus. 
The hamiltonian of a free proton is 
$\hat{H}_0$ $= \sqrt{\hat{M}^2 + {\bf q}^2}$,
and the full hamiltonian of the struck proton in the medium is
\begin{equation}
  \hat{H} = \hat{H}_0 + \hat{V} 
  = \sqrt{\hat{M}^2 + {\bf q}^2} + \hat{V}.
\label{HF3}
\end{equation}
We then introduce a transverse-size dependent potential 
due to soft interaction for the proton travelling in the nuclear medium 
\cite{Low:PRD}:
\begin{equation}
  \hat{V} = -{\it i} c_0 \; 
           ( \hat{{\bf r}}_{\perp}^2 + \hat{{\bf s}}_{\perp}^2 ), 
\end{equation}
where ${\bf r}_\perp$ and ${\bf s}_\perp$ 
are the transverse size of the proton, 
and $c_0$ is a constant.

In the present situation, 
the recoil momentum of the proton is much larger 
than the proton mass, and thus 
we use the ultra-relativistic approximation 
for the full hamiltonian, $\hat{H}$, {\it i.e.}, 
\begin{equation}
  \hat{H} \simeq |{\bf q}| + \frac{\hat{M}_{\rm I}^2}{2 |{\bf q}|}, 
\label{HF5}
\end{equation}
where
\begin{equation}
   -\hat{M}_{\rm I}^2 
   = \eta \left( \hat{p}_{r}^{2} + \hat{p}_{s}^{2} 
      + \alpha^2 (\hat{r}_{0}^2 - \hat{r}_{3}^2 
                + \hat{s}_{0}^2 - \hat{s}_{3}^2) 
      + \alpha_{\perp}^2 (\hat{{\bf r}}_{\perp}^2 
                        + \hat{{\bf s}}_{\perp}^2)  \right) + C.
\label{HF5a}
\end{equation}
Here we have defined a ``transverse" size parameter 
\begin{equation}
  \alpha_{\perp} 
  \equiv \alpha \times 
   \sqrt{1 - {\it i} \; {2 |{\bf q}| c_0 \over \eta \alpha^2}}.
\end{equation}
The eigenfunction, $\phi_{{\rm I}, n}(r, s; {\bf q})$, 
obeys the following wave equation
\begin{equation}
  \hat{M}_{{\rm I}}^{2} \; |\phi_{{\rm I}, n}; {\bf q} \rangle 
=    M_{{\rm I}, n}^{2} \; |\phi_{{\rm I}, n}; {\bf q} \rangle. 
\end{equation}
The wave function and the mass eigenvalue are 
\begin{eqnarray}
   \phi_{{\rm I}, n}(r, s; {\bf q})
       &=& N_{0}^2 N_{n_{1}}^{\prime} N_{n_{2}}^{\prime} 
       \cdots N_{n_{6}}  \\
    &{}& \times
       \exp \{ {\alpha \over 2} (r_{0}^{2} - r_{3}^{2} 
                + s_{0}^{2} - s_{3}^{2} 
               - {2 \over M_{n}^{2}} (q' \cdot r)^{2} 
               - {2 \over M_{n}^{2}} (q' \cdot s)^{2}) \} \nonumber\\
    &{}& \times
           \exp \{ -{\alpha_{\perp} \over 2} 
                    ({\bf r}_{\perp}^{2} + {\bf s}_{\perp}^{2}) \} 
                    \nonumber\\ 
    &{}& \times
            H_{n_{1}}(\sqrt{2\alpha_{\perp}}\; r_{1})  \;
            H_{n_{2}}(\sqrt{2\alpha_{\perp}}\; r_{2})  \;
            H_{n_{3}} \left( \sqrt{2\alpha} 
               \sqrt{{1 \over M_{n}^{2}} (q' \cdot r)^{2} -
        (r_{0}^{2} - r_{3}^{2}) } \right)  \nonumber\\
    &{}& \times
            H_{n_{4}}(\sqrt{2\alpha_{\perp}}\; s_{1})  \;
            H_{n_{5}}(\sqrt{2\alpha_{\perp}}\; s_{2})  \;
            H_{n_{6}} \left( \sqrt{2\alpha} 
               \sqrt{{1 \over M_{n}^{2}} (q' \cdot s)^{2} 
                    - (s_{0}^{2} - s_{3}^{2}) } \right) , \nonumber
\label{eq:rqm14}
\end{eqnarray}
where $q_{\mu}'$ $= (M + q_{0}, {\bf 0}_{\perp}, q_3)$,  
$N_{0}^2$ $= \sqrt{\alpha} / \sqrt{\pi}$, 
$N_{n}^{2}$ $= \sqrt{\alpha} / (\sqrt{\pi} n!)$, 
$N_{n}^{\prime 2}$ $= \sqrt{\alpha_{\perp}} / (\sqrt{\pi} n!)$,  
and
\begin{eqnarray}
   M_{{\rm I}, n}^{2} &=& \eta \{-(2 n_{r, 0} + 1) \alpha 
          + \left( 2(n_{1} + n_{2}) + 2 \right) \alpha_{\perp}
          + ( 2 n_{3} + 1) \alpha  \nonumber \\ 
         &{}& - (2 n_{s, 0} + 1) \alpha 
            + \left(2(n_{4} + n_{5}) + 2 \right) \alpha_{\perp}
                  + (2 n_{6} + 1) \alpha \} - C.  
\label{eq:rqm16}
\end{eqnarray}
$n_{i}$ $(i = 1, \cdots, 6)$ are non-negative integers, 
and $n_{r, 0} = n_{s, 0} = 0$. 
Here also the proton is assumed to run in the $z$-direction. 
$H_n(x)$ are the Hermite polynomials defined 
in Appendix~\ref{sec-hermite}.

\subsection{Longitudinal-Transverse Correlation}
\label{sec-ltcor}

The model in its original form has a problem 
in describing the color transparency.  
The hard scattering operator, 
\begin{eqnarray}
\hat{O}(q) &=& \exp \{ {\it i} q \cdot (x_{1} - X) \}
            = \exp \{- {\it i} \sqrt{2} q \cdot s \} 
\end{eqnarray}
cannot excite the transverse modes.  
We have assumed that the quark, 1, has been struck. 
Since there is no correlation between 
the longitudinal motion and the transverse motion in the original model, 
the transverse modes remain in their ground state, 
which is fatal in describing the color transparency.

The absence of the longitudinal-transverse correlation is 
a special feature of the harmonic oscillator model. 
The correlation exists in any other form of the potential. 
For example, Frankfurt {\it et al.} \cite{Frankfurt1} 
showed that several models did contain such a correlation. 
We discuss the case of the Coulomb model in Appendix~\ref{sec-coulomb}, 
and leave more elaborate discussions to Ref.~\cite{Mak:PRC}. 
In order to remedy the defect of the model, 
we modify the operator, $\hat{O}(q)$, 
so as to incorporate the correlation in the ground state. 
We thus introduce a new parameter, $\nu$, 
representing the strength of the correlation
and replace $\hat{O}(q)$ by 
\begin{equation}
  \hat{O}(q) = \exp \{- {\it i} \sqrt{2} q \cdot s \} 
  \times
  \exp \{ \nu \; q^2 (\hat{{\bf r}}_{\perp}^2 
                                 + \hat{{\bf s}}_{\perp}^2) \}, ~~~
              (q^2 < 0).  
\end{equation}
We have incorporated not only the longitudinal-transverse correlation, 
but also the correlation among the quarks in $\hat{O}(q)$. 

Then, the modified form factor, $F_{{\rm ep}}^{\nu}(q^2)$, becomes,
\begin{eqnarray}
  F_{{\rm ep}}^{\nu}(q^2) 
  &\equiv&  \langle \phi_0; {\bf q}|
  e^{ - {\it i} \sqrt{2} q \cdot \hat{s}} \;
  e^{\nu  q^2 (\hat{{\bf r}}_{\perp}^2 + \hat{{\bf s}}_{\perp}^2)} 
          | \phi_0; {\bf 0} \rangle \nonumber\\
 &=& \left( 1 + \nu \; {-q^2 \over\alpha} \right)^{-2} \;
     \left( 1 + \frac{-q^2}{2 M_0^2}  \right)^{-2} \;
  \times
  \exp \left( -\frac{1}{2 \alpha}\; 
            \frac{-q^2}{ 1 + (-q^2)/(2 M_0^2) } \right) 
            \nonumber \\ 
 &=& \left( 1 + \nu \; {-q^2 \over\alpha} \right)^{-2} \;
      \times
           F_{{\rm ep}}(q^2). 
\label{form2}
\end{eqnarray}
It is well known that the original form of the form factor, 
{\it i.e.}, the case of $\nu = 0$ of eq.~(\ref{form2}), 
gives an excellent description of the observed 
proton charge form factor \cite{Namiki1}. 
The introduction of $\nu$ somewhat spoils the beauty 
and actually gives an undesirable
$Q^{-8}$ behavior at large $Q^2$ $(= -q^2)$. 
However, for modest values of $\nu$, 
we can choose the size parameter, $\alpha$, 
so as to obtain a reasonable fit to the observed form factor 
in the relevant region of $Q^2$.

\subsection{Proton Survival Amplitude}
\label{sec-amp}

We are now in the position to calculate the proton survival amplitude, 
$M^{(D)}_{{\rm eA}}(q^2; t)$, in eq.~(\ref{matrix0}),
which can be used to obtain the proton survival probability 
through eq.~(\ref{P-}). 
Inserting the complete set of the eigenstates of $\hat{H}$
into the matrix element, we have
\begin{eqnarray}
  M^{(D)}_{{\rm eA}}(q^2; t)
 &=& \langle \phi_0; {\bf q}| e^{- {\it i} \hat{H} t} \; 
     e^{ - {\it i} \sqrt{2} q \cdot s} \;
     e^{ \nu q^2 ( \hat{{\bf r}}_{\perp}^2  
                 + \hat{{\bf s}}_{\perp}^2 )} 
         |\phi_0; {\bf 0} \rangle  \nonumber \\
 &=&  \sum_{n = 0}^{\infty} 
         \langle \phi_0; {\bf q}| e^{- {\it i} \hat{H} t} 
              | \phi_{{\rm I}, n}; {\bf q} \rangle 
         \langle \phi_{{\rm I}, n}; {\bf q} |
     e^{ - {\it i} \sqrt{2} q \cdot s} \;
     e^{ \nu q^2 ( \hat{{\bf r}}_{\perp}^2  
                 + \hat{{\bf s}}_{\perp}^2 )} 
         |\phi_0; {\bf 0} \rangle. 
\label{matrix-c1}
\end{eqnarray} 
The sum over $n$ in the above expression is rather complicated 
due to the $M_{n}$-dependence of the wave function, 
$|\phi_{{\rm I}, n}; {\bf q} \rangle$. 
Although it could be calculated numerically since the convergence is fast, 
we use here an approximation of replacing $M_{n}$ by $M_{0}$ 
in order to get a closed expression for the survival amplitude. 
The approximation amounts to taking the same velocity 
for all the intermediate states in the sum. 
It is justified since the $M_{n}$-dependence is not very strong and 
the large-$n$ states do not contribute significantly to the sum. 

Each matrix element is given as follows. 
The first one is 
\begin{eqnarray}
  &{}& \langle \phi_0; {\bf q}| e^{- {\it i} \hat{H} t} 
            |\phi_{{\rm I}, n}; {\bf q} \rangle   
   = e^{- {\it i} \{ |{\bf q}| 
         + \eta M_{{\rm I}, n}^2 / (2 |{\bf q}|) \} t } \;
   \int d^{4}r d^{4}s \; \phi_0(r, s; {\bf q})  
                         \phi_{{\rm I}, n}(r, s; {\bf q})   \nonumber \\
 &=& \delta_{n_3, 0} \delta_{n_6, 0} \times 
     \exp \left[- {\it i} \left\{ |{\bf q}| + \frac{\eta}{|{\bf q}|}
     \left( (n_1 + n_2 + n_4 + n_5) \alpha_{\perp} 
    + 2 \alpha_{\perp} - C/2 \right) \right\}  t \right] \nonumber \\
 &{}& \times
     \frac{4 \alpha \alpha_{\perp}}{(\alpha_{\perp} + \alpha)^2} \;
     \frac{\sqrt{n_1! n_2! n_4! n_5!}}
          {\left( n_{1}/2 \right)! \left( n_{2}/2 \right)!
           \left( n_{4}/2 \right)! \left( n_{5}/2 \right)!} \;
   \left( \frac{1}{2} \; 
          \frac{\alpha_{\perp} - \alpha}{\alpha_{\perp} + \alpha} 
   \right)^{(n_1 + n_2 + n_4 + n_5)/2}, 
\label{l-matrix}
\end{eqnarray}
where $n_1$, $n_2$, $n_4$, and $n_5$ are even.  
We have used eq.~(\ref{eq:herm7}) in Appendix~\ref{sec-hermite}. 
Another one is 
\begin{eqnarray}
 &{}& \langle \phi_{{\rm I}, n}; {\bf q}|
   e^{- {\it i} \sqrt{2} q \cdot  s} \;
   e^{\nu q^2 (\hat{{\bf r}}_{\perp}^2 
             + \hat{{\bf s}}_{\perp}^2)}  |\phi_0; {\bf 0} \rangle  
 =  \int d^{4}r d^{4}s \;  \phi_{{\rm I}, n}(r, s; {\bf q}) 
   e^{- {\it i} \sqrt{2} q \cdot  s} \;
   e^{\nu q^2 (\hat{{\bf r}}_{\perp}^2 
             + \hat{{\bf s}}_{\perp}^2)}  \phi_0(r, s; {\bf 0})  
       \nonumber \\
  &=& \delta_{n_3, 0} \times 
\frac{4 \alpha \alpha_{\perp}}
           {(\alpha + \alpha_{\perp} + 2 \nu (-q^2))^{2}} \;
  \left( -{\it i} \frac{q_3}{\sqrt{{\alpha}}} 
                  \frac{1}{1 + (q_0/M_0) } \right)^{n_6} 
          \nonumber\\
  &{}& \times
   \frac{\sqrt{n_1! n_2! n_4! n_5!}}
        {\left( n_{1}/2 \right)! \left( n_{2}/2 \right)!
         \left( n_{4}/2 \right)! \left( n_{5}/2 \right)!} 
      \times
   \left( \frac{1}{2} 
          \frac{\alpha_{\perp} - \alpha - 2 \nu (-q^2)} 
               {\alpha_{\perp} + \alpha + 2 \nu (-q^2)} 
   \right)^{(n_1 + n_2 + n_4 + n_5)/2} \nonumber \\
  &{}& \times
  \left( 1 + \frac{-q^2}{2 M_0^2} \right)^{-2} \;
    \exp \left( - \frac{1}{2 \alpha} 
                  \frac{-q^2}{1 + (-q^2)/(2 M_0^2)} 
         \right), 
\label{r-matrix}
\end{eqnarray}
where $n_1$, $n_2$, $n_4$, and $n_5$ are even. 
We have used eq.~(\ref{eq:herm10}) in Appendix~\ref{sec-hermite}.

Summing up the products of these matrix elements, 
eqs.~(\ref{l-matrix}) and (\ref{r-matrix}),
we obtain a closed form for the survival amplitude, 
eq.~(\ref{matrix-c1}), 
\begin{eqnarray}
   M^{(D)}_{{\rm eA}}(q^2; t)
   &=& \sum_{n = 0}^{\infty}
       \langle \phi_0; {\bf q}| e^{- {\it i} \hat{H} t} 
       | \phi_{{\rm I}, n}; {\bf q} \rangle
       \langle \phi_{{\rm I}, n}; {\bf q} |
        e^{- {\it i} \sqrt{2} q \cdot  s} \;
        e^{\nu q^2 (\hat{{\bf r}}_{\perp}^2 
                  + \hat{{\bf s}}_{\perp}^2)} 
       |\phi_0; {\bf 0} \rangle  \nonumber \\
   &=& \left( \frac{4 \alpha \alpha_{\perp}}
            {(\alpha + \alpha_{\perp} + 2 \nu (-q^2))
             (\alpha_{\perp} + \alpha)} \right)^2 \;
    \exp \left[- {\it i} \left\{ |{\bf q}| + \frac{\eta}{|{\bf q}|}
                          (2 \alpha_{\perp} + C/2) \right\}  t 
         \right] \nonumber \\
   & & \times
       \left[ 1 - 
         \left( \frac{\alpha_{\perp} - \alpha - 2 \nu (-q^2)}
                     {\alpha_{\perp} + \alpha + 2 \nu (-q^2)}
         \right) \;
         \left( \frac{\alpha_{\perp} - \alpha}{\alpha_{\perp} + \alpha} 
         \right) \;
     \exp \{ - {\it i} \frac{\eta}{|{\bf q}|}  \;
                       2 \alpha_{\perp}  t \} \right]^{-2} \nonumber  \\
   & & \times  F_{{\rm ep}}(q^2). 
\label{matrix-c2}
\end{eqnarray}
Here we have used a formula,
\begin{equation}
  \sum_{m = 0}^{\infty} {(2m - 1)!! \over m!} \; (2 x)^{m}
   = (1 - 4 x)^{-1/2}.
\end{equation}
The squared amplitude is given by
\begin{eqnarray}
   |M^{(D)}_{{\rm eA}}(q^2; t)|^2
   &=& | \langle \phi_0; {\bf q}| e^{- {\it i} \hat{H} t} \;
        e^{- {\it i} \sqrt{2} q \cdot  s} \;
        e^{\nu q^2 (\hat{{\bf r}}_{\perp}^2 
                  + \hat{{\bf s}}_{\perp}^2)} 
       |\phi_0; {\bf 0} \rangle |^2  \nonumber \\
  &=&  \left( F_{{\rm ep}}(q^2) \right)^2 \; 
         \left| \frac{4 \alpha \alpha_{\perp}}
            {(\alpha + \alpha_{\perp} + 2 \nu (-q^2))
             (\alpha_{\perp} + \alpha)} \right|^4  \; \nonumber\\
  &{}& \times 
       \left| 1- 
         \left( \frac{\alpha_{\perp} - \alpha - 2 \nu (-q^2)}
                     {\alpha_{\perp} + \alpha + 2 \nu (-q^2)}
         \right) \;
         \left( \frac{\alpha_{\perp} - \alpha}{\alpha_{\perp} + \alpha} 
         \right) \;
         \exp \{ - {\it i} \frac{\eta}{|{\bf q}|} \; 
                       2 \alpha_{\perp}  t \} \right|^{-4} \nonumber  \\
   & & \times
       \left| \exp \{ - {\it i} \frac{\eta}{|{\bf q}|} \;
                        2 \alpha_{\perp}  t  \} \right|^2. 
\label{matrix-c3}
\end{eqnarray}
Finally, we obtain the ratio, $R(q^2; t)$, of eq.~(\ref{ratio1}) as
\begin{eqnarray}
   R(q^2; t) 
   &=& \left| 1 - 
       \left( \frac{\alpha_{\perp} - \alpha - 2 \nu (-q^2)}
                   {\alpha_{\perp} + \alpha + 2 \nu (-q^2)}
       \right) \;
       \left( \frac{\alpha_{\perp} - \alpha}{\alpha_{\perp} + \alpha} 
       \right) \;
       \right|^{4} \;   \nonumber \\
   & & \times
       \left| 1 - 
       \left( \frac{\alpha_{\perp} - \alpha - 2 \nu (-q^2)}
                   {\alpha_{\perp} + \alpha + 2 \nu (-q^2)}
       \right) \;
       \left( \frac{\alpha_{\perp} - \alpha}{\alpha_{\perp} + \alpha} 
       \right) \;
       \exp \{ - {\it i} \frac{\eta}{|{\bf q}|} \; 
                       2 \alpha_{\perp}  t \} \right|^{-4} \nonumber  \\
   & & \times
       \left| \exp \{ - {\it i} \frac{\eta}{|{\bf q}|} \; 
                       2 \alpha_{\perp}  t \} \right|^{2}.
\label{P-c}
\end{eqnarray}
Given the survival probability, $P^{(-)}({\bf q}; {\bf r})$ 
$= R(q^2; t({\bf r}))$, as eq.~(\ref{P-}), 
we can calculate the nuclear transparency using eq.~(\ref{T-P}).

Let us examine whether $R(q^2; t)$ and therefore $T(q)$ 
become unity when $Q^2$ $\rightarrow \infty$ 
in our model in accord with the idea of the color transparency. 
The closed expression allows us to examine the large-$Q^2$ behavior
of the survival probability. 
>From the definition of the transverse size parameter, $\alpha_{\perp}$, 
the exponents of eq.~(\ref{P-c}) become unity, {\it i.e.},  
\begin{equation}
        \exp \{ - {\it i} \frac{\eta}{|{\bf q}|} \; 
                       2 \alpha_{\perp}  t \} 
      \rightarrow 1, ~~( |{\bf q}| \rightarrow \infty). 
\end{equation}
Thus, the survival probability 
becomes unity in the large-$Q^2$ limit.

\newpage
\setcounter{equation}{0}
\section{Numerical Results and Discussions}
\label{sec-nume}

In this section we present our numerical results and discussions 
after we fix the parameters. 

\subsection{Parameters}
\label{sec-para}

With the modified expression for the form factor,
we determine the parameter, $\alpha$, 
so as to reproduce the observed form factor 
in the relevant region of $Q^2$. 
Specifically, we require 
$F_{{\rm ep}}^{\nu}(q^2)$ $= 7.11 \times 10^{-3}$ 
at $Q^2$ $(= -q^2)$ $= 7.512 ~[({\rm GeV/c})^2]$ \cite{Cow:PRL}. 
Using this value of $\alpha$, 
we show the form factor, eq.~(\ref{form2}), in Fig.~1
with the observed one. 
We see that the observed charge form factor of the proton 
is reasonably well reproduced by the modified  
version as long as $\nu$ is not too large.

The parameters, $\eta$ and $C$,
are adjusted to reproduce the proton mass 
of 940~[MeV] and the Roper resonance of 1440~[MeV]. 
We have assumed that the Roper resonance corresponds to 
a second excited state:
\begin{eqnarray}
  4 \eta \alpha - C = M_{p}^2, ~~~
  8 \eta \alpha - C = M_{{\rm Roper}}^2.
\end{eqnarray}
>From these relations, we obtain the value of $C$ and 
a relation of $\eta \alpha$ as follows; 
\begin{eqnarray}
  C           &=& M_{{\rm Roper}}^2 - 2 M_{p}^2 ~~~ 
              = 7.90 \;[{\rm fm}^{-2}], \\
  \eta \alpha &=& \frac{1}{4}(M_{{\rm Roper}}^2 - M_{p}^2) 
              = 7.66 \;[{\rm fm}^{-2}].
\end{eqnarray}
The numbers of $\alpha$ and $\eta$ for each fixed $\nu$ 
are given in Table~\ref{table1}.

\begin{table}
\begin{center}
\begin{tabular}{|c|cc|}  \hline
  $\nu$ & $\alpha$~ $[{\rm fm}^{-2}]$ & $\eta$  \\
  \hline  
  0.00  &  11.3                       & 0.68  \\
  0.02  &  15.5                       & 0.49   \\
  0.05  &  21.1                       & 0.36   \\
  0.25  &  53.7                       & 0.14    \\ \hline
\end{tabular}
\end{center}
\caption{Parameters, $\alpha$ and $\eta$, for each fixed $\nu$. }
\label{table1}
\end{table}

We then determine the parameter, $c_0$, 
which represents the strength of the absorptive interaction, 
$\hat{V}$, using the relation, eq.~(\ref{c-inert}), 
as discussed in sec.~\ref{sec-timedep}. 
We use the linearly interpolated values 
of the experimental proton-proton reaction cross section for 
$\sigma_{{\rm NN}}^r(|{\bf q}|)$ \cite{LB}, 
and substitute $\rho_{{\rm NM}}$ $= 0.17 \;[{\rm fm}^{-3}]$ 
in the calculation of Figs.~2-4 and in Table~\ref{table2}. 
In calculating the nuclear transparency in Figs.~5-10, 
we take $\rho_{{\rm NM}}$ $\simeq (A - 1) \rho_0$, 
referring to the comments given in connection with 
eq.~(\ref{c-inert}).

In the present model, the relation, eq.~(\ref{c-inert}), becomes
\begin{eqnarray}
  -2 {\rm Im}~\langle \hat{V} \rangle 
  & = &  2 c_0 \langle \phi_0; {\bf q}| 
                (\hat{{\bf r}}^2_\perp 
              +  \hat{{\bf s}}^2_\perp) |\phi_0; {\bf q} \rangle
\nonumber \\
 & = & 4 \; \frac{c_0}{\alpha} 
   =   \sigma_{{\rm NN}}^{r}(|{\bf q}|) \; \rho_{0} \; v_q. 
\label{c-d0}
\end{eqnarray}
$c_0$ is thus determined as 
\begin{equation}
  c_0 = {1 \over 4} \; \alpha \; 
        \sigma_{{\rm NN}}^r (|{\bf q}|) \; \rho_{0} \; v_q. 
\end{equation}
The numbers of $c_0$ for each fixed $\nu$ 
are given in the Table~\ref{table2}.

\begin{table}
\begin{tabular}{|c|r|r|r|r|r|r|}                       \hline
$Q^{2}~[({\rm GeV/c})^{2}]$ & $q_{3}$~[GeV] & $\sigma_{{\rm NN}}^{r}$~[mb] 
  & \multicolumn{4}{c|}{$c_{0}$ ~[${\rm fm}^{-3}$]}  \\        \hline
  {} &      {}&     {}& $\nu = 0.00$ & $\nu = 0.02$ 
                         & $\nu = 0.05$ & $\nu = 0.25$\\  \cline{4-7}
   1 &  1.133 &  7.52 & 0.278 & 0.382 & 0.519 & 1.322 \\
   2 &  1.771 & 23.36 & 0.991 & 1.360 & 1.851 & 4.711 \\
   4 &  2.923 & 27.04 & 1.236 & 1.696 & 2.309 & 5.876 \\
  10 &  6.197 & 28.16 & 1.337 & 1.834 & 2.496 & 6.353 \\
  20 & 11.558 & 29.70 & 1.422 & 1.950 & 2.655 & 6.757 \\
 100 & 54.220 & 30.85 & 1.481 & 2.032 & 2.766 & 7.040 \\ \hline
\end{tabular}
\caption{Absorption strength, $c_0$, for each fixed $\nu$. }
\label{table2}
\end{table}

\subsection{Survival Probability}

The numerical results for the ratio, $R(q^2; t)$, in eq.~(\ref{P-c})
with the parameters determined in the previous subsection 
are shown in Figs.~2-4. 
The difference between the curve for the inert proton 
and those for the dynamical proton is considered to reflect 
the effect of the internal dynamics.
We can expect the color transparency effect 
to start already at $Q^2$ $(= -q^2)$ $= 4 \;[({\rm GeV/c})^2]$, 
and to become prominent above $Q^2$ $= 10 \;[({\rm GeV/c})^2]$.
The internal dynamics gives rise to a fluctuating absorption, 
and effectively weakens its strength.
That is, the internal dynamics could cause the color transparency,  
and the relativity plays an important role there. 

Here we comment on the case of $\nu = 0$ 
as an interesting comparison. 
As was discussed in subsec.~\ref{sec-ltcor}, 
the hard interaction operator, $\hat{O}(q)$, does not 
excite the transverse modes for $\nu = 0$. 
One might thus expect that this would correspond to the inert case. 
Actually, however, the transverse modes are excited 
by the interaction, $\hat{V}$, 
as the proton travels through nuclear medium, 
though the difference between the curves for $\nu = 0$ 
and the inert proton is small. 
The survival probability for $\nu = 0$ is 
that for the proton in its ground state at $t = 0$ 
and is a quantity relevant in discussing proton-nucleus 
elastic and inelastic scatterings. 
This is another interesting subject which can be treated 
in the present model, 
but will not be discussed further in this paper.

\subsection{Nuclear Transparency}

We numerically calculate the nuclear transparency 
for several target nuclei by the formulation in sec.~\ref{sec-timedep}. 
We parametrize the nuclear densities 
of $^{12}$C, $^{56}$Fe, and $^{197}$Au as the Woods-Saxon form, 
\begin{equation}
  \rho({\bf r}) = \rho_0 \; {1 \over 1 + e^{(r - c)/z}}, ~~~
  \int d{\bf r} \; \rho({\bf r}) = 1, 
\end{equation}
where $c$ $= 1.1 \times A^{1/3}$ ~[fm] and $z$ $= 0.53$~[fm]. 
The numerical results are shown in Fig.~5 as a function of $q^2$, 
and in Figs.~6-10 as a function of the target mass number ($A$). 
The $A$-dependence is included through the parameter, $c$, above. 
To include the weak $q^2$-dependence of the proton-nucleon 
reaction cross section, 
we use the spline function for fitting the experimental data 
of the reaction cross section \cite{LB}. 
We see that the internal dynamics enhances the nuclear transparency 
as expected in Figs.~5-10. 

>From the results shown in Fig.~5, 
one would expect that the increase of the nuclear transparency 
due to the internal dynamics could be observed already 
at $Q^2$ $(= -q^2)$ $= 4~[({\rm GeV/c})^2]$. 
This appears to contradict with the recent experiment at SLAC 
\cite{Makins}, \cite{ONe:PRL}, 
though the cases of weak longitudinal-transverse correlation 
have not been excluded due to its large errors. 
Slow increase of the transparency is still permitted. 

We further calculate the target mass number dependence 
($A$-dependence) of the nuclear transparency in the present model. 
The situation turns out to be more encouraging. 
The numerical results are shown in Figs.~6-10. 
In Figs.~6-8 we plot our results with the experimental data for 
$Q^2$ = 3.06, 5.00, 6.77 ~[$({\rm GeV/c})^2$], respectively 
\cite{Makins} \cite{ONe:PRL}. 
The asymptotic line of the inert proton 
is well fitted by 1.62 $\times A^{-1/3}$. 
As one can see from Figs.~6 and 7, 
the data are almost explained by the inert proton, 
and the data points for heavy targets are on the line of $A^{-1/3}$, 
while in Fig.~8 a deviation is observed for $^{197}$Au.  
We plot our predictions for larger $Q^2$ in Figs.9 and 10,   
and the deviations are prominent for these cases. 

The importance of the $A$-dependence as a probe of 
the color transparency was previously pointed out 
by Ralston \cite{Ral:Pre}, 
and by Kohama, Yazaki and Seki \cite{Koh:PL1}. 
In Ref.~\cite{Koh:PL1}
it was found that deviations from $A^{-1/3}$-dependence 
for large-$A$ regime could be a signature of the color transparency 
based on a {\em model-independent} discussion. 
We call it the ``off-$A^{-1/3}$ criterion". 
According to the off-$A^{-1/3}$ criterion, 
the data point for $^{197}$Au at $Q^2$ = 6.77 ~[$({\rm GeV/c})^2$]
can be a signature of the color transparency, 
though the error is large. 
This data point is unique in that it deviates 
to the above of the 
asymptotic $A^{-1/3}$ line of the inert case. 
In Ref.~\cite{ONe:PRL} they fitted the data 
with the effective cross sections 
instead of the reaction cross section in the Glauber expression 
emphasizing the small-$A$ data, 
though the $A^{-1/3}$-dependence is universal only for large-$A$. 
Therefore we suggest that the measurement for 
heavy nuclei should be done with high precision, 
and see whether the data lie on the line of $A^{-1/3}$ or not.  
If the data are significantly off the line, 
it may signal the onset of the color transparency. 
The Jefferson Lab. is one of the best place to carry out 
this kind of experiments.  

The relevant parameters in our model are $\eta$, $\alpha$, and $\nu$. 
For a given $\nu$, $\alpha$ is determined 
to reproduce the charge form factor.  
$\nu$ is thus the crucial parameter, 
and it may be that we should take it smaller than 0.02 
which is the smallest in our choices. 
If this is the case, 
and if the longitudinal-transverse correlation is weak, 
then the $(e, e'p)$ reaction may not be the best process 
in observing the color transparency. 

In the case of $(p, 2p)$, for example, 
since the incident proton and the scattered proton propagate 
in the directions different from that of the momentum transfer, 
the color transparency might be more clearly observed for them, 
though the situation is essentially the same as that 
for the $(e, e'p)$ reaction for the recoil proton. 
The calculation of the nuclear transparency 
for the $(p, 2p)$ reaction is desirable, 
and will be done in near future.

\newpage
\setcounter{equation}{0}
\section{Summary and Conclusions} 
\label{sec-summary}

We have constructed a relativistic quantum-mechanical model 
to discuss the effects of the internal structure of the proton 
on the nuclear transparency.
The model describes the proton as a three-quark system 
bound by a four-dimensional harmonic oscillator potential. 
It reproduces both the electro-magnetic form factors of the proton 
and the overall features of its excitation spectrum 
represented by a linearly rising Chew-Frautschi trajectory very well. 

We have modified the hard interaction operator, $\hat{O}(q)$, 
so as to incorporate the longitudinal-transverse correlation 
which is crucial in discussing the color transparency for $(e, e'p)$, 
but is absent in the original model. 

The most important feature of the model is that
the relativistic effects in the internal dynamics 
are properly taken into account 
and it allows us the correct treatment of the large-momentum 
transfer limit, {\it i.e.}, $Q^2$ $(= -q^2)$ $\rightarrow \infty$. 

The final-state interaction of the dynamical proton 
with the nuclear medium is taken to be purely absorptive, 
and the strength is assumed to be proportional to 
the transverse squared-size of the travelling proton. 
With the model specified in this way, 
we have calculated the proton survival probability 
in the nuclear medium after it is struck by the incident electron, 
and used the results to estimate the nuclear transparency. 
For comparison, we have made similar calculations for the inert proton. 

The effects of the internal dynamics are clearly observed, 
and are found to become more and more important 
as the momentum transfer squared, $Q^2$, increases.  
The nuclear transparency goes to unity as $Q^2 \rightarrow \infty$. 

The present calculation is intended to discuss qualitative features of 
the color transparency in a specific model for the internal 
structure of the proton. 
The calculation indicates an early onset of the color transparency, 
which appears to contradict with the recent experiment at SLAC, 
though the cases of weak longitudinal-transverse correlation 
have not been excluded due to the large error bars of the data. 
The situation looks more encouraging, 
if we see the data from the $A$-dependence. 
We have to refine the choice of the parameters to make the model 
more quantitative. 
The strength of the longitudinal-transverse correlation is crucial 
in this respect, 
and thus the nuclear transparency of the $(e, e'p)$ 
offers a way of studying such details of the internal dynamics.

The hadronic process, such as $(p, 2p)$, on the other hand, 
may be more suitable for observing the color transparency itself, 
since the incident and the scattered proton can 
get transverse momentum transfers. 
The calculation of the nuclear transparency for such processes 
in our model will be an interesting subject for the future.

\begin{center}
{\bf Acknowledgement} 
\end{center}

We would like to thank all the members of the nuclear theory group
at University of Tokyo for their assistance
and encouragement during the course of this work.
A.K. and K.Y. would like to show their gratitude 
to Prof.~N.Isgur and to Prof.~F.Lenz for the discussions. 

This work is supported by a Grand-Aid 
of the Japanese Ministry of Education, 
Science and Culture at the University of Tokyo (No.~08640355).

\newpage
\appendix
\begin{center}
  {\bf APPENDIX}
\end{center}

\setcounter{section}{0}
\setcounter{equation}{0}
\section{Hermite Polynomials}
\label{sec-hermite}

In this Appendix we show the definition of the Hermite polynomials, 
and and derive some useful formulae. 

We define the Hermite polynomial in terms 
of the generating function as 
\begin{equation}
  e^{t x - t^{2}/2} = \sum_{n = 0}^{\infty} 
                       {t^n \over n!} H_{n}(x).
\label{eq:herm1}
\end{equation}
Note that there are two different definitions for this functions. 
The factor two on the exponential is different. 
The orthogonality is 
\begin{equation}
  \int_{-\infty}^{\infty} dx \;
           e^{- x^{2}/2} \; H_{n}(x) H_{m}(x) 
           = \delta_{n, m} \; n! \; \sqrt{2 \pi}. 
\label{eq:herm2}
\end{equation}

Let us consider the one-dimensional harmonic oscillator model. 
The hamiltonian is given by 
\begin{eqnarray}
  H &=& - \frac{1}{2 m} \frac{d^2}{dx^2} 
       + {1 \over 2} m \omega^2 x^2 \nonumber\\
    &=& \frac{\omega}{2} \left( - \frac{d^2}{d\xi^2} + \xi^2 \right), 
   ~~~  \xi = \alpha x,   
\label{eq:herm3}
\end{eqnarray}
where $\xi$ is a dimensionless variable, 
and $\alpha^2$ $= m \omega$. 
The Schr\"{o}dinger equation is given by 
\begin{equation}
 H \phi_{n}(x) = E_n \phi_{n}(x), 
\label{eq:herm4}
\end{equation}
and the wave function is 
\begin{equation}
  \phi_{n}(\xi) = N_n \; e^{- \xi^{2}/2} \; H_{n}(\sqrt{2} \xi), ~~~ 
    N_{n}^2 = {\alpha \over \sqrt{\pi} n! }, 
\label{eq:herm5}
\end{equation}
and the energy is 
\begin{equation}
  E_n = \left( n + {1 \over 2} \right), ~~(n = 0, 1, 2, \cdots). 
\label{eq:herm6}
\end{equation}

We then derive two formulae which are 
convenient to calculate matrix elements in subsec.~\ref{sec-amp}. 
The first formula is 
\begin{equation}
  \int_{-\infty}^{\infty} d\xi \;
           e^{- \gamma_0 \xi^{2}/2} \; H_{2m}(\xi)  
   = \sqrt{{\pi \over \gamma_0}} \; 
           \frac{(2m)!}{m!} \; 
           \left({1 \over 4 \gamma_0} - {1 \over 2} \right)^m, ~~~
      (m = 0, 1, 2, \cdots),  
\label{eq:herm7}
\end{equation}
where $\gamma_0$ is complex. 
To prove eq.~(\ref{eq:herm7}), 
we use the generating function, eq.~(\ref{eq:herm1}), as 
\begin{equation}
  \int_{-\infty}^{\infty} d\xi \; 
    \exp \{- \gamma_0 \xi^2 + t x - {t^{2} \over 2} \} 
  = \sum_{n = 0}^{\infty} {t^n \over n!} \;
    \int_{-\infty}^{\infty} d\xi \; 
         e^{- \gamma_0 \xi^2 } \; H_{n}(\xi).
\label{eq:herm8}
\end{equation}
The l.h.s. of eq.~(\ref{eq:herm8}) gives 
\begin{eqnarray}
   \int_{-\infty}^{\infty} d\eta \; 
           \exp \{- \gamma_0 \eta^2 + t \xi - {t^{2} \over 2} \} 
   &=& \sqrt{{\pi \over \gamma_0}} \; 
       \exp \{ \left({1 \over 4 \gamma_0} - {1 \over 2} \right) t^2 \}
   \nonumber\\
   &=& \sqrt{{\pi \over \gamma_0}} \;
   \sum_{m = 0}^{\infty} {1 \over m!} \;
           \left({1 \over 4 \gamma_0} - {1 \over 2} \right)^m 
           t^{2m}, 
\label{eq:herm9}
\end{eqnarray}
where $m$ $= 0, 1, 2, \cdots$. 
Comparing eq.~(\ref{eq:herm9}) with 
the r.h.s. of eq.~(\ref{eq:herm8}), 
we obtain eq.~(\ref{eq:herm7}).

The second formula is 
\begin{eqnarray}
  &{}& \int_{-\infty}^{\infty} d\xi \;
           e^{- s_0 \xi^{2} + {\it i} \kappa \xi} \; H_{n}(\beta \xi)  
\nonumber \\
   &=& 
 \left\{ \begin{array}{l} 
   \sqrt{\pi / s_0} \; e^{- \kappa^2/(4 s_0)} \;
       \left( \sqrt{1 - {\beta^2 \over 2 s_0}} \right)^n \;
       H_n \left( {{\it i} \beta \kappa \over 
                     2 s_0 \sqrt{1 - \beta^2 / (2 s_0) }} \right), 
       ~({\rm if}~  \beta^2 \neq 2 s_0)  \\
      \sqrt{\pi / s_0} \; e^{- \kappa^2/(4 s_0)} \;
       \left( {\it i} \beta \kappa / (2 s_0) \right)^n, 
       ~({\rm if}~  \beta^2 = 2 s_0).
\end{array} \right. 
\label{eq:herm10}
\end{eqnarray}
%Note that $H_n({\it i} x)$ $= {\it i}^n \; H_n(x)$. 
To prove this expression, we can apply the same method 
as that for the first one. 
We then write 
\begin{equation}
  \int_{-\infty}^{\infty} d\xi \; 
    \exp \{- s_0 \xi^2 + {\it i} \kappa \xi
             + \beta t \xi - {t^{2} \over 2} \} 
  = \sum_{n = 0}^{\infty} {t^n \over n!} \;
    \int_{-\infty}^{\infty} d\xi \; 
         e^{- s_0 \xi^2 + {\it i} \kappa \xi} \; H_{n}(\beta \xi).
\label{eq:herm11}
\end{equation}
The l.h.s. of eq.~(\ref{eq:herm11}) gives
\begin{eqnarray}
  &{}& \int_{-\infty}^{\infty} d\xi \; 
    \exp \{- s_0 \xi^2 + {\it i} \kappa \xi
             + \beta t \xi - {t^{2} \over 2} \} \nonumber\\
   &=& \sqrt{{\pi \over s_0}} \; 
       \exp \left( - \frac{\kappa^2}{4 s_0} \right) 
     \times 
       \exp \{ -\frac{1}{2} 
                  \left( 1 - {\beta^2 \over 2 s_0} \right) t^2 
                + {1 \over 2 s_0} {\it i} \beta t \kappa \} 
    \nonumber \\
   &=& \sqrt{{\pi \over s_0}} \; 
       \exp \left( - \frac{\kappa^2}{4 s_0} \right)  
    \times 
       \sum_{n = 0}^{\infty} {t^n \over n!} \;
       \left( \sqrt{1 - {\beta^2 \over 2 s_0}} \right)^n \; 
        H_{n} \left( \frac{{\it i} \beta \kappa}
                         {2 s_0 \sqrt{1 - (\beta^2 / 2 s_0)}} 
              \right).  
\label{eq:herm12}
\end{eqnarray}
If $\beta^2$ $= 2 s_0$, 
then the l.h.s. of eq.~(\ref{eq:herm11}) gives
\begin{eqnarray}
  \int_{-\infty}^{\infty} d\xi \; 
    \exp \{- s_0 \xi^2 + {\it i} \kappa \xi
            + \beta t \xi - {t^{2} \over 2} \} 
   &=& \sqrt{{\pi \over s_0}} \; 
       \exp \left( - \frac{\kappa^2}{4 s_0} \right)  \;
       \exp \{ {{\it i} \beta \kappa \over 2 s_0} t \} 
    \nonumber \\
   &=& \sqrt{{\pi \over s_0}} \; 
       \exp \left( - \frac{\kappa^2}{4 s_0} \right) \;
       \sum_{n = 0}^{\infty} {t^n \over n!} \;
       \left({ {\it i} \beta \kappa \over 2 s_0} \right)^n.  
\label{eq:herm13}
\end{eqnarray}
If we compare eqs.~(\ref{eq:herm12}) and (\ref{eq:herm13}) 
with the r.h.s. of eq.~(\ref{eq:herm11}), 
we obtain eq.~(\ref{eq:herm10}).

\newpage
\setcounter{equation}{0}
\section{Case of the Coulomb Potential}
\label{sec-coulomb}

In subsec.~\ref{sec-ltcor}, we have modified the hard interaction operator, 
$\hat{O}(q)$, so as to incorporate the longitudinal-transverse 
correlation which is crucial in discussing the color transparency, 
but is absent in the harmonic oscillator model. 
The correlation strength is represented by the parameter, $\nu$, 
but we have no {\it a priori} basis for determining the strength. 
Here, we use a non-relativistic Coulomb model to get a rough 
idea of the strength.

In the Coulomb model, the ground state is given by
\begin{equation}
  \phi_{c, 0}(r) = 2 \alpha_c^{3/2} \; e^{-\alpha_c r}.
\label{coulomb-gstate}
\end{equation}
Here, $\alpha_c$ is determined as
\begin{eqnarray}
 \langle {\bf r}^2 \rangle
 &\equiv& \langle \phi_{c, 0} | {\bf r}^2 | \phi_{c, 0} \rangle
    = 3 / \alpha_c^2 = (0.853)^2 ~[{\rm fm}^2], \nonumber \\ 
 &\Leftrightarrow& 
       \alpha_c^2 = \frac{3}{\langle {\bf r}^2 \rangle}.  
\label{a_c-d}
\end{eqnarray}
We use $\langle {\bf r}^2 \rangle$ $= (0.853)^2 ~[{\rm fm}^{2}]$.

We introduce the following quantity to see the correlation strength. 
It becomes unity in the absence of correlation. 
\begin{equation}
   \left( {2 \over 3} \;\langle {\bf r}^2 \rangle \right)^{-1}
   \times
     \frac{\langle \phi_{c, 0}| (x^2 + y^2) \; 
                    e^{{\it i} q z} | \phi_{c, 0} \rangle}
          {\langle \phi_{c, 0}| e^{{\it i} q z} | \phi_{c, 0}\rangle}
      = \left( 1 + {|{\bf q}|^2 \over 4 \alpha_c^2} \right)^{-1}, 
\label{coulomb-t}
\end{equation}
where the first term is introduced for the normalization. 
We can see that the correlation effect is appreciable  
at a large-momentum transfers.

We now calculate the same quantity 
in a non-relativistic harmonic oscillator model
with $\nu$ being introduced in a way analogous to the relativistic model:
\begin{equation}
   \left( {2 \over 3} \;\langle {\bf r}^2 \rangle \right)^{-1}
   \times
   \frac{\langle \phi_{h.o., 0} | (x^2 + y^2) 
            e^{- \nu q^2 (x^2 + y^2)} \; e^{{\it i} q z} 
           | \phi_{h.o., 0} \rangle}
        {\langle \phi_{h.o., 0} | 
            e^{- \nu q^2 (x^2 + y^2)} \; e^{{\it i} q z} 
           | \phi_{h.o., 0} \rangle}
  = \left(1 + {\nu |{\bf q}|^2 \over \alpha_{h.o.} } \right)^{-1}.
\label{oscillator-t}
\end{equation}
$\alpha_0$ is determined as
\begin{eqnarray}
 \langle {\bf r}^2 \rangle
 &\equiv& \langle \phi_{h.o., 0} | {\bf r}^2 | \phi_{h.o., 0} \rangle
    = 3 / \alpha_{h.o.}, \nonumber \\ 
 &\Leftrightarrow& \alpha_{h.o.} = \frac{3}{\langle {\bf r}^2 \rangle}.
\label{a_0-d}
\end{eqnarray}

Thus, $\nu = 1/4$ for the Coulomb model. 
It seems that among the commonly-used binding potentials, 
the Coulomb potential has the strongest correlation. 
We take this value for $\nu$ as the maximum 
in our calculation for the nuclear transparency.

\newpage
\begin{center}
{\Large Figure Captions}
\end{center}
\mbox{}\\
{\large Figure 1}\\
{\normalsize A comparison of the electromagnetic form factors, 
$F_{{\rm ep}}^{\nu}(q^2)$, of the proton in the present model 
with experimental data.
The dotted curve is the form factor of the original harmonic oscillator 
model ($\nu$ = 0). 
The solid curve, the dashed curve, and the dot-dashed curve 
are those of the modified harmonic oscillator model 
with $\nu$ = 0.02, 0.05, 0.25, respectively. 
The crosses and the circles indicate 
the experimental data from Ref.~\cite{Cow:PRL}.
}\\
\mbox{}\\
{\large Figure 2}\\
{\normalsize The time development of the ratio, 
$|R(q^2; t)|^2$, eq.~(\ref{P-c}),  
for $Q^{2}$ $(= -q^{2})$ = 10~ [$({\rm GeV/c})^{2}$].
The dotted curve is the case of the inert proton. 
The solid curve, the dashed curve, and the dot-dashed curve 
are those of the dynamical proton 
with $\nu$ = 0.02, 0.05, 0.25, respectively. 
}\\
\mbox{}\\
{\large Figure 3}\\
{\normalsize The time development of the ratio, 
$|R(q^2; t)|^2$, eq.~(\ref{P-c}),  
for $Q^{2}$ $(= -q^{2})$ = 20~ [$({\rm GeV/c})^{2}$].
The curves are the same as those of Fig.~3.
}\\
\mbox{}\\
{\large Figure 4}\\
{\normalsize The time development of the ratio, 
$|R(q^2; t)|^2$, eq.~(\ref{P-c}),  
for $Q^{2}$ $(= -q^{2})$ = 100~ [$({\rm GeV/c})^{2}$].
The curves are the same as those of Fig.~3.
}\\
\mbox{}\\
{\large Figure 5}\\
{\normalsize A comparison of the nuclear transparency 
with experimental data for $^{12}$C, $^{56}$Fe, and $^{197}$Au 
as a function of $Q^2$ $(= -q^{2})$. 
The dotted curve is the case of the inert proton. 
The solid, dashed, and dot-dashed curves are 
the case of the dynamical proton 
with $\nu$ = 0.02, 0.05, 0.25, respectively. 
The data are from Ref.~\cite{ONe:PRL}. 
}\\
\mbox{}\\
{\large Figure 6}\\
{\normalsize A comparison of the nuclear transparency 
with experimental data for $^{12}$C, $^{56}$Fe, and $^{197}$Au 
as a function of the target mass number, $A$. 
$Q^2$ $(= -q^{2})$ = 3.06 ~[$({\rm GeV/c})^2$]. 
The dotted curve is the case of the inert proton. 
The solid, dashed, and dot-dashed curves are 
the case of the dynamical proton 
with $\nu$ = 0.02, 0.05, 0.25, respectively. 
The data are from Ref.~\cite{ONe:PRL}. 
}\\
\mbox{}\\
{\large Figure 7}\\
{\normalsize A comparison of the nuclear transparency 
with experimental data for $^{12}$C, $^{56}$Fe, and $^{197}$Au 
as a function of the target mass number, $A$. 
$Q^2$ $(= -q^{2})$ = 5.00 ~[$({\rm GeV/c})^2$]. 
The curves are the same as those of Fig.~6.
}\\
\mbox{}\\
{\large Figure 8}\\
{\normalsize A comparison of the nuclear transparency 
with experimental data for $^{12}$C, $^{56}$Fe, and $^{197}$Au 
as a function of the target mass number, $A$. 
$Q^2$ $(= -q^{2})$ = 6.77 ~[$({\rm GeV/c})^2$]. 
The curves are the same as those of Fig.~6.
}\\
\mbox{}\\
{\large Figure 9}\\
{\normalsize The nuclear transparency 
as a function of the target mass number, $A$. 
$Q^2$ $(= -q^{2})$ = 10.0 ~[$({\rm GeV/c})^2$]. 
The curves are the same as those of Fig.~6.
}\\
\mbox{}\\
{\large Figure 10}\\
{\normalsize The nuclear transparency 
as a function of the target mass number, $A$. 
$Q^2$ $(= -q^{2})$ = 20.0 ~[$({\rm GeV/c})^2$]. 
The curves are the same as those of Fig.~6.
}

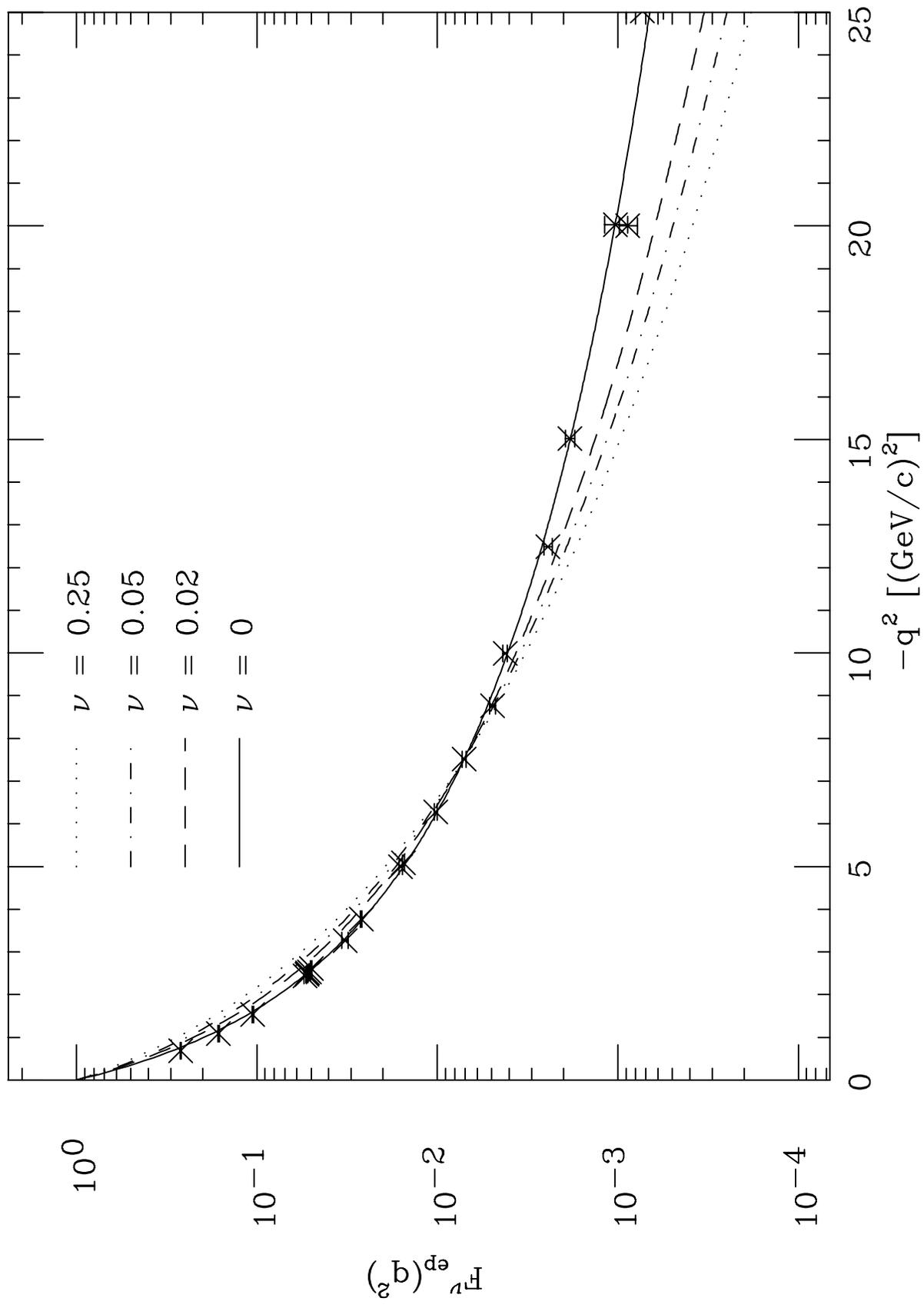
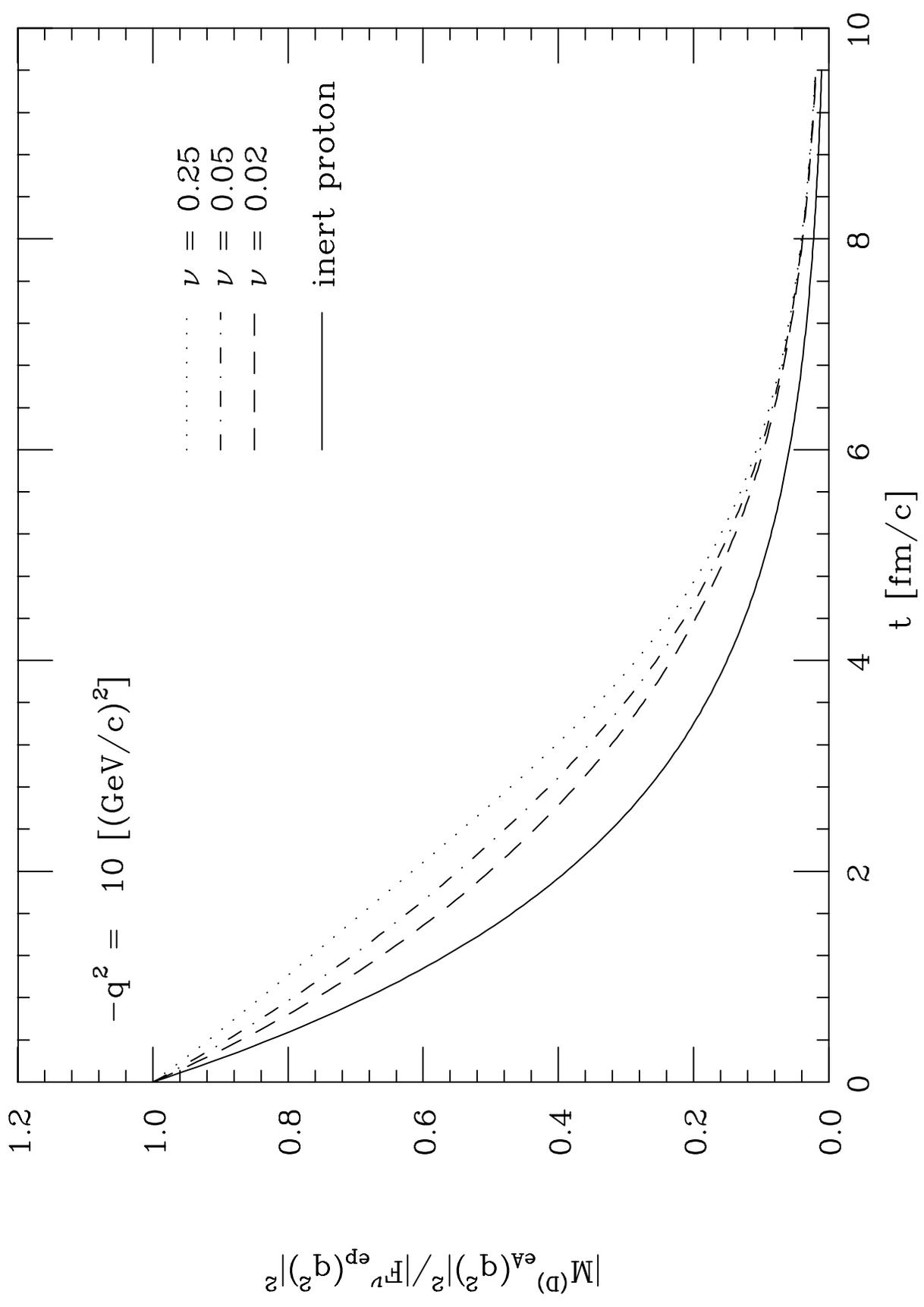
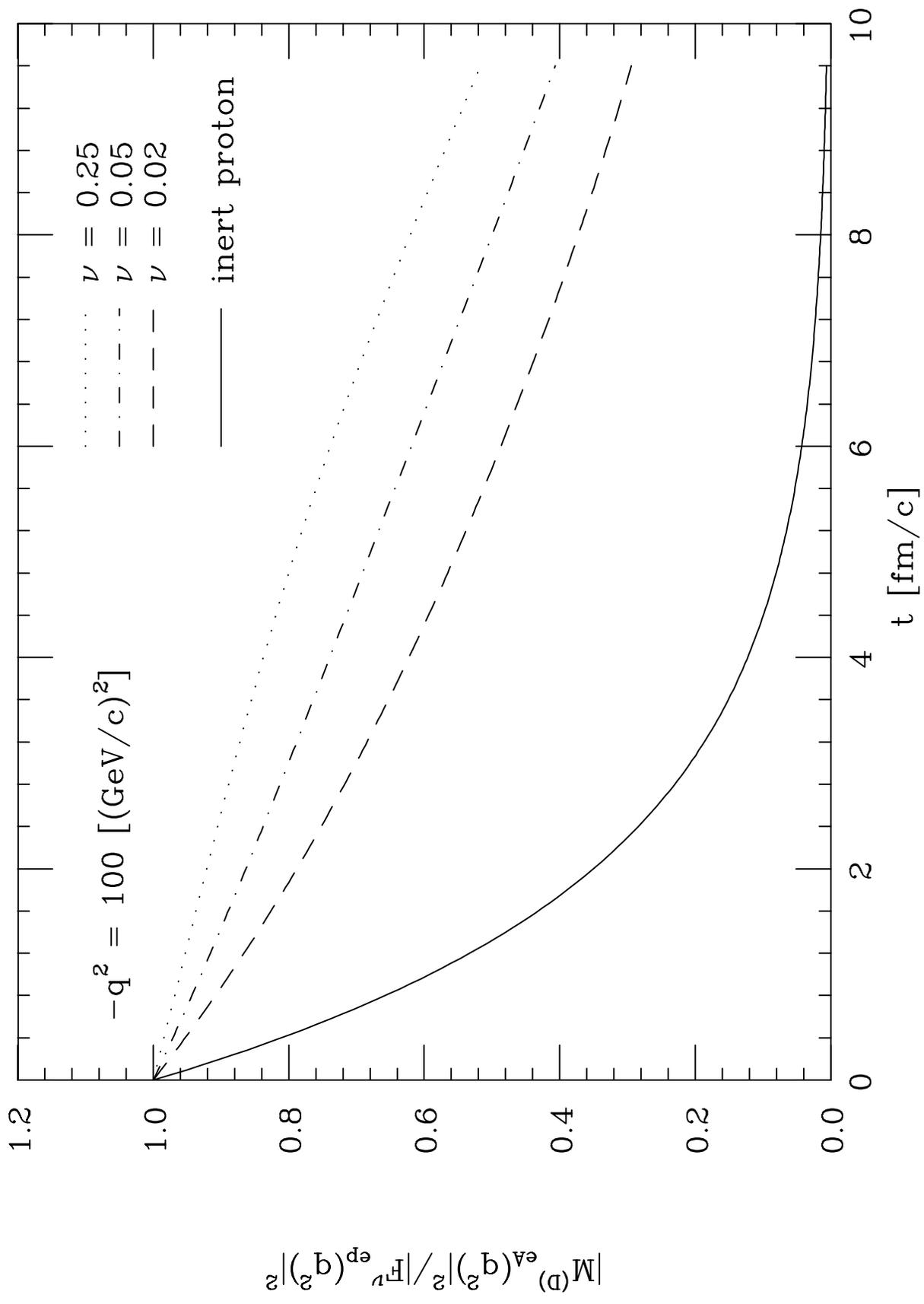
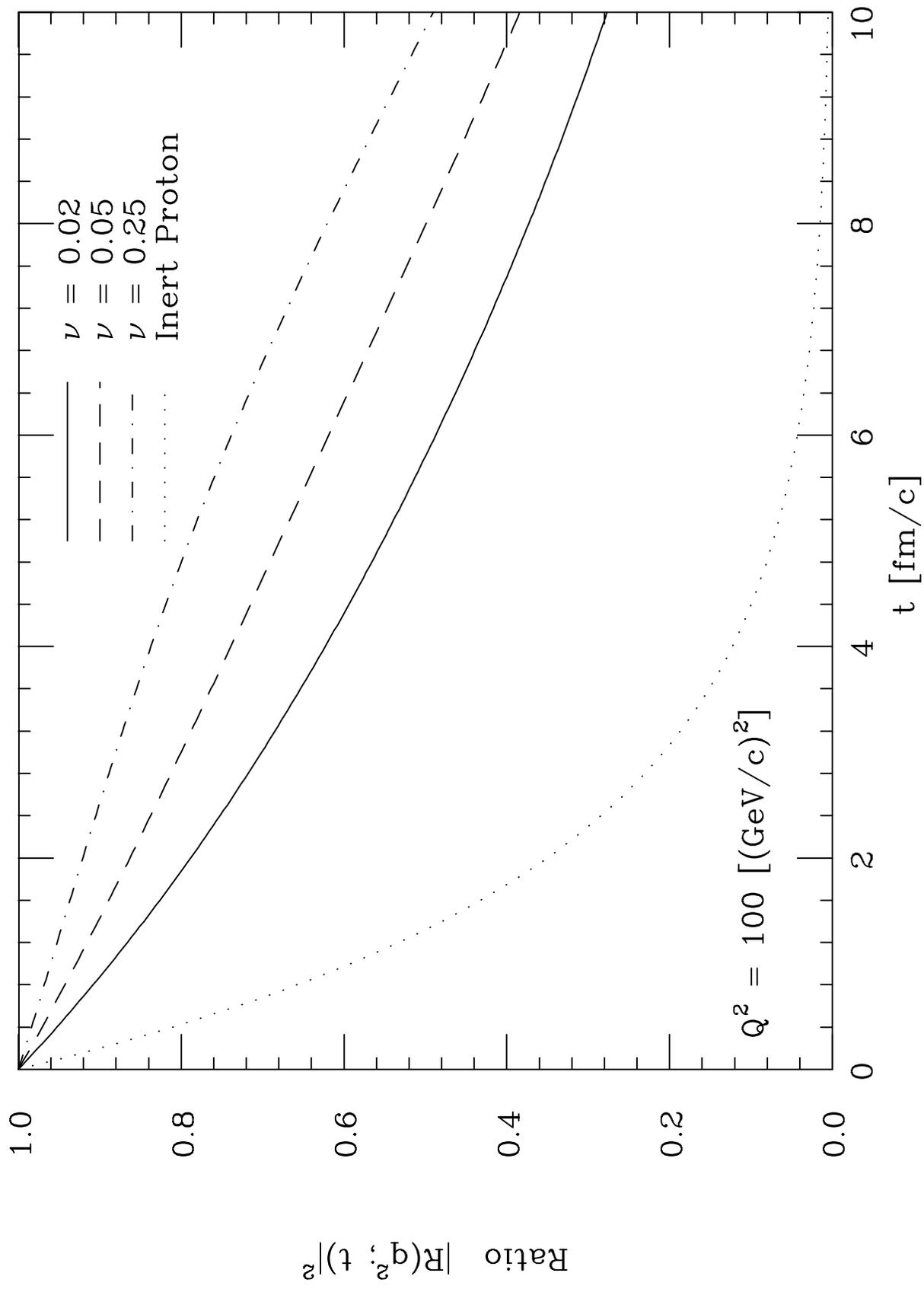
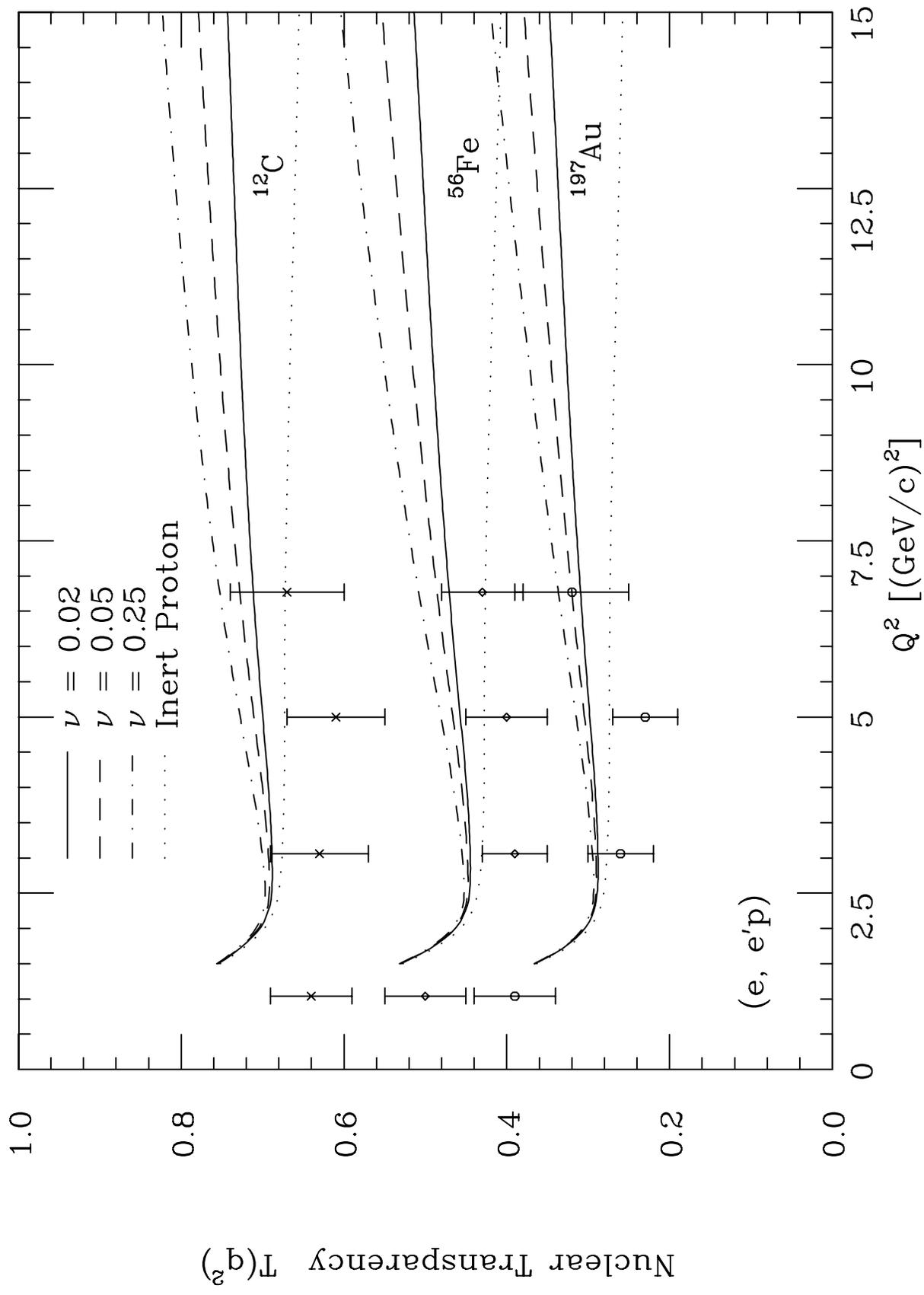
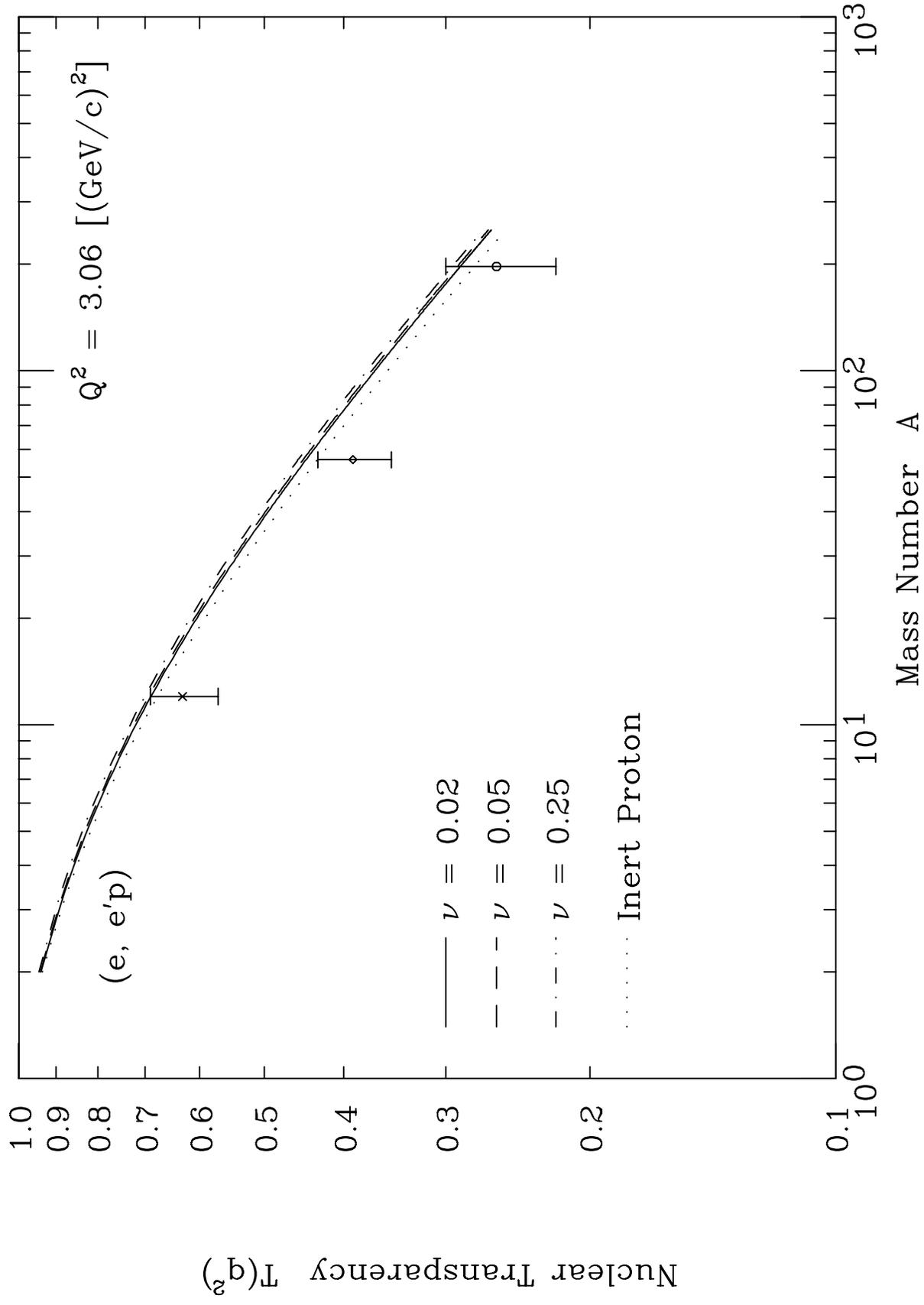
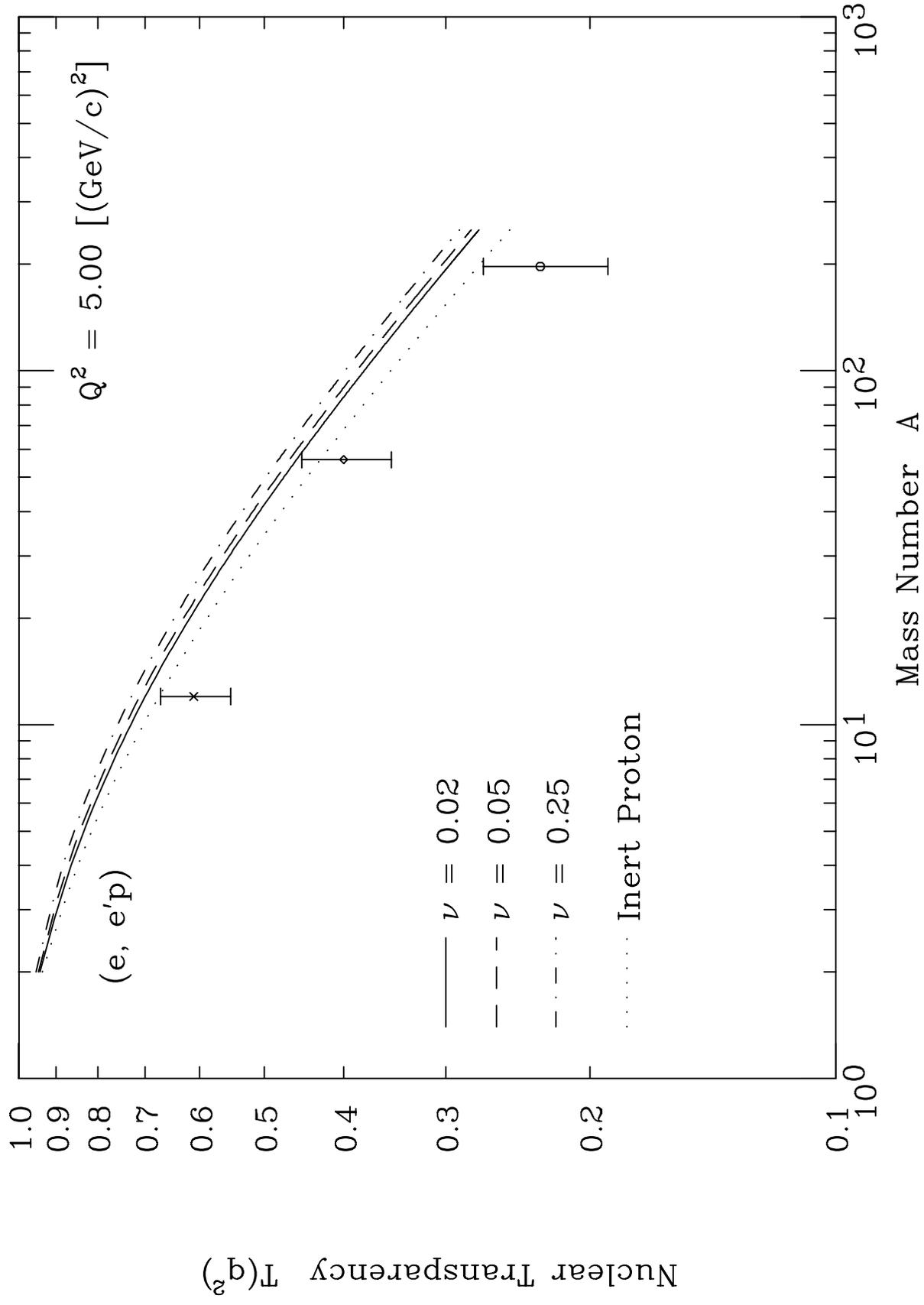
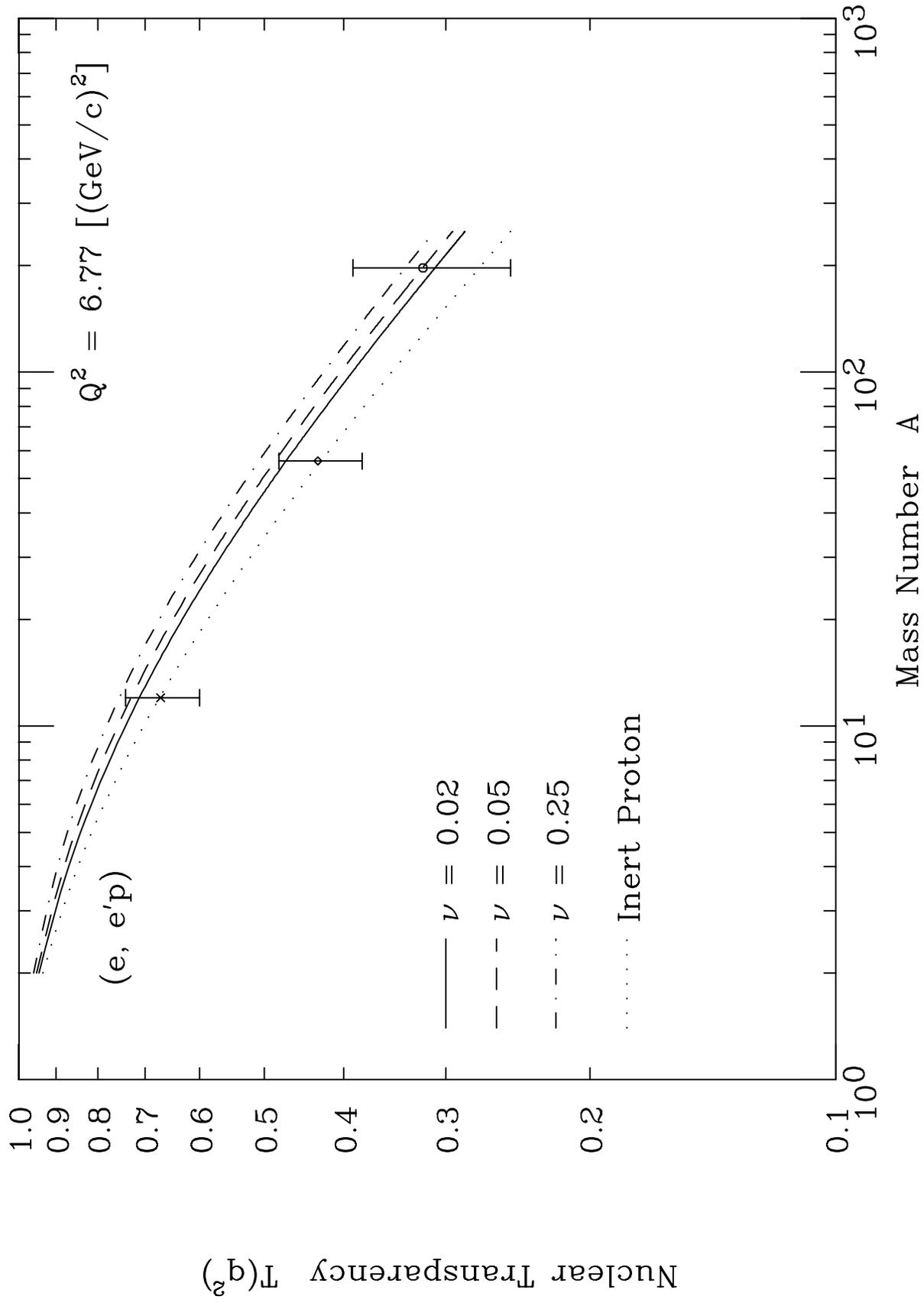
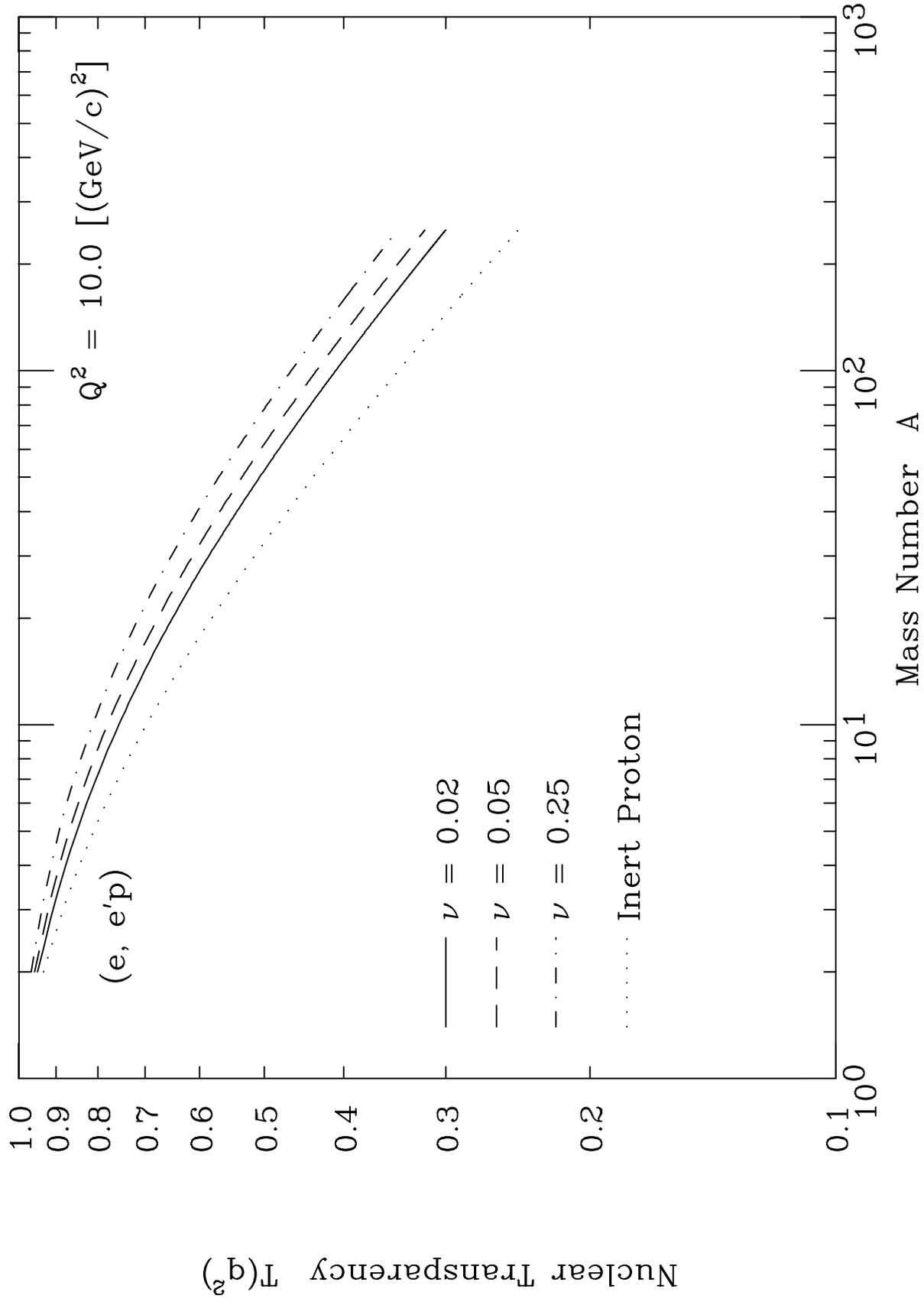
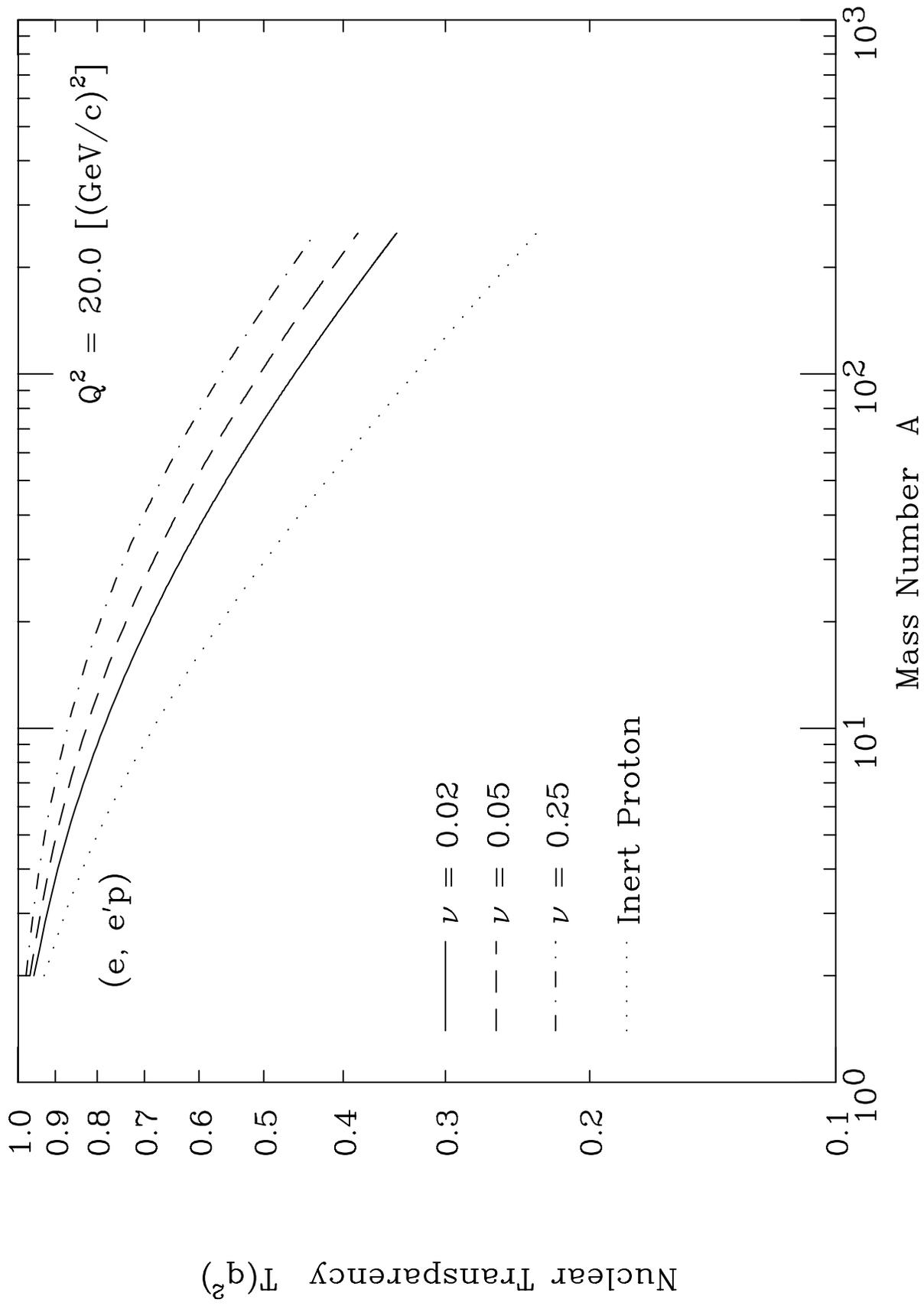
\end{document}